\documentclass[twocolumn,epjc3]{svjour3}
\smartqed  
\RequirePackage{fix-cm}
\RequirePackage[T1]{fontenc}
\RequirePackage{graphicx}
\RequirePackage{mathptmx}      
\RequirePackage{flushend}
\RequirePackage[numbers,sort&compress]{natbib}
\RequirePackage[colorlinks,citecolor=blue,urlcolor=blue,linkcolor=blue]{hyperref}
\RequirePackage{latexsym}
\RequirePackage{multirow}
\RequirePackage{amsmath,amssymb,mathtools}
\RequirePackage{caption}
\RequirePackage{subcaption}
\RequirePackage{gensymb,textcomp}

\graphicspath{ {Plots/} }

\RequirePackage{hyphenat}
\DeclareMathOperator{\erfc}{erfc}
\DeclareMathOperator{\erf}{erf}

\journalname{Eur. Phys. J. C}
\begin{document}
\title{Reconstruction of air shower muon lateral distribution functions using integrator and binary modes of underground muon detectors}
\author{V.V Kizakke Covilakam\thanksref{e1,addr1,addr2}
	 \and 
	 A.D. Supanitsky\thanksref{addr1}
	 \and 
	 D. Ravignani\thanksref{addr1}
}
\thankstext{e1}{e-mail: varada.varma@iteda.cnea.gov.ar(corresponding author)}
\institute{Instituto de Tecnolog\'{\i}as en Detecci\'{o}n y Astropart\'{\i}culas (CNEA, CONICET, UNSAM), Av. General Paz 1499, San Mart\'{\i}n, B1650KNA, Buenos Aires, Argentina\label{addr1}
	\and
	Karlsruher Institut f\"{u}r Technologie, Institut f\"{u}r Kernphysik, Hermann-von-Helmholtz-Platz 1, Eggenstein-Leopoldshafen, 76344, Baden-Württemberg, Germany.
 \label{addr2}
}
\date{Received: date / Accepted: date}
	\maketitle
\begin{abstract}
     The investigation of cosmic rays holds significant importance in the realm of particle physics, enabling us to expand our understanding beyond atomic confines. However, the origin and characteristics of ultra-high-energy cosmic rays remain elusive, making them a crucial topic of exploration in the field of astroparticle physics. Currently, our examination of these cosmic rays relies on studying the extensive air showers (EAS) generated as they interact with atmospheric nuclei during their passage through Earth's atmosphere. Accurate comprehension of cosmic ray composition is vital in determining their source. Notably, the muon content of EAS and the atmospheric depth of the shower maximum serve as the most significant indicators of primary mass composition. In this study, we present two novel methods for reconstructing particle densities based on muon counts obtained from underground muon detectors (UMDs) at varying distances to the shower axis. Our methods were analyzed using Monte Carlo air shower simulations. To demonstrate these techniques, we utilized the muon content measurements from the UMD of the Pierre Auger cosmic ray Observatory, an array of detectors dedicated to measuring extensive air showers. Our newly developed reconstruction methods, employed with two distinct UMD data acquisition modes, showcased minimal bias and standard deviation. Furthermore, we conducted a comparative analysis of our approaches against previously established methodologies documented in existing literature.
\end{abstract}
\section{Introduction}\label{intro}
    \paragraph{} Cosmic rays constitute a population of highly energetic elementary particles and nuclei with an unknown origin that descend upon Earth from outer space. Their spectrum follows a nearly power law distribution, spanning from approximately 10$^9$ eV to 10$^{20}$ eV \cite{Mollerach:17}. Direct measurement of primary cosmic rays with sufficient flux, which occurs at low energies, is feasible through experiments conducted in space. Nevertheless, for energies surpassing approximately 10$^{15}$ eV, the flux weakens, necessitating reliance on the interactions between primary particles and atmospheric molecules to generate secondary particles called extensive air showers (EAS) \cite{Mathews:05}. These showers can be observed during their progression in the atmosphere, either on the Earth's surface or underground. The Pierre Auger Observatory, positioned in the southern hemisphere \cite{Auger:20}, encompasses detectors capable of investigating cosmic ray showers at all three levels: during their development in the atmosphere, on the surface, and underground. Consequently, these showers are reconstructed to examine the primary particles' three principal observables: energy spectrum, arrival direction, and chemical composition. Notably, the Pierre Auger Observatory employs Underground Muon Detectors (UMD) to directly measure the muon content of the showers. High-energy muons exhibit superior penetration capabilities compared to other secondary particles. Subterranean experiments have demonstrated that the density of muons serve as indications of the primary cosmic ray nuclei's chemical composition and energy spectrum. However, this sensitivity is constrained by a threshold imposed by the thickness of the soil covering. Muons possess a unique sensitivity to composition due to a phenomenon where lighter particles (e.g., protons) exhibit lower efficiency in producing multiple muons when compared to heavier nuclei. Although underground detectors cannot determine the energy and specific type of primary particles on an event-to-event basis, it is possible to derive information about the mass composition by comparing the measured distributions of muon multiplicities with those calculated through precise Monte Carlo simulations employing trial models of the primary spectrum and composition.
	\paragraph{} The UMD at the Pierre Auger Observatory measures the fall in muon density with increasing distance to the shower axis, known as the muon lateral distribution function (MLDF) \cite{Brian:15}. Once completed, the UMD will be equipped with a total of 73 muon detectors. Among these detectors, a triangular array with a spacing of 750 m will encompass approximately 20 km$^2$ of the area, housing 61 muon detectors. The remaining 12 detectors will be arranged in a smaller triangular array with a spacing of 433 m, covering approximately 1 km$^2$. For the purpose of this study, we will solely focus on the 750 m array. These detectors within the 750 m array possess the capability to measure showers with energies ranging from around $10^{16.5}$ eV to $10^{19}$ eV and can detect events up to a zenith angle of 45$\degree$. Each grid location is equipped with three 10 m$^2$ modules, segmented into 64 plastic scintillation strips containing embedded wavelength-shifting optical fibers that are coupled to an array of 64 silicon photomultipliers (SiPMs) \cite{AMIGA:21}. Collectively, these three modules form a 30 m$^2$ detector with 192 segments. Data acquisition within the UMD is triggered by the associated Surface Detector (SD) station. To minimize contamination from energetic electrons and gamma particles accompanying muons near the shower core, the detectors are buried at a depth of 2.3 m underground. The soil density at the UMD site measures 540 g cm$^{-2}$, corresponding to a shielding equivalent to 22 radiation lengths. This ensures absorption of the electromagnetic component of the Extensive Air Shower (EAS), enabling only muons with energies above approximately $\sim 1$ GeV$/\cos\theta_\mu$ (where $\theta_\mu$ denotes the zenith angle of muon motion) to reach the buried detectors.
	\paragraph{} The UMD encompasses two distinct acquisition modes, specifically the binary mode and the integrator mode, also referred to as the Analog to Digital Converter (ADC) mode. These modes operate  simultaneously to provide a broad range of detection capabilities \cite{AMIGA:21}. Each mode employs distinct methodologies to transform raw signals into the number of muons.
	\paragraph{} The counting of muons is accomplished by the binary mode segment-wise, which tallies signals exceeding a certain amplitude threshold \cite{Flavia:22}. Each of the 64 SiPM signals is processed independently by a pre-amplifier, a fast shaper, and a discriminator located in every channel of two 32-channel ASICs \cite{Eng:21}. Subsequently, the signal from the discriminator undergoes sampling into 64 2048-bit traces with a 3.125 ns sampling time, accomplished via a Field-Programmable Gate Array (FPGA). The binary mode has a total inhibition window of 37.5 ns (12 samples $\times$ 3.125 ns) \cite{icrc:19}. This method of muon counting is independent of the gain or fluctuation of the optical device, is independent of the muon impact position, and does not require a thick scintillator. However, it tends to count two muons that arrive simultaneously on the same strip as one because of the pile-up effect. This statistical error can be corrected as long as the number of strips with simultaneous signals is lower than the total segmentation \cite{Flavia:22}. The binary mode saturates if the majority of strips in a given module simultaneously receive a signal, and it can only detect a limited number of muons at the same time, which limits its ability to probe at distances close to the shower axis. The accuracy of the detector resolution in binary mode is also limited by geometric biases such as the corner clipping bias. Overcounting occurs when a muon passes through the lateral edge of a scintillator bar and is detected in both bars, producing the corner clipping bias \cite{Juan:17}. As more particles hit the detector, the resolution of the binary mode decreases, and the densities closer to the shower cannot be accurately determined.
	\paragraph{} The ADC mode determines the charge of all segments and converts them to the number of muons. This is done by dividing the total signal charge by the average charge of a single vertical muon signal \cite{Eng:20}. During ADC acquisition mode, the 64 SiPM signals are added up analogically, and the resulting sum is amplified using low-gain and high-gain amplifiers with an amplification factor ratio of about 4. These two channels are called Low Gain (LG) and High Gain (HG) channels, respectively. The signals are sampled every 6.25 ns (twice the binary sampling period). The fluctuations in the ADC mode reduce with the number of detected muons. However, as it depends on the estimation of particles from an integrated signal, signal fluctuations propagate to uncertainties in the estimation of the number of muons. The ADC mode has a resolution of 60$\%$ for single-muon signals, which decreases inversely with the square root of the number of particles. It reaches 10$\%$ after several tens of muons, matching the resolution of the binary mode. With increasing muons, the ADC mode provides much better resolution than the binary mode, allowing for better precision in measuring particle densities closer to the core \cite{AMIGA:21}. The operating range of the LG and HG channels in the ADC mode overlaps with the binary mode and depends on the muon density \cite{Eng:20}.
	\paragraph{} Another experiment that measures the muon content of the air shower is the Akeno Air Shower Observatory. It consisted of a 100 km$^2$ air shower array called the Akeno Giant Air Shower Array (AGASA) \cite{AGASA1}. The experiment completed operations in 2004 and later merged with the High Resolution Fly’s Eye (HiRes) group, forming the Telescope Array Collaboration \cite{AGASA2}. AGASA has played a vital role in providing critical information about cosmic rays. Surface and muon detectors with concrete shielding were installed in AGASA. AGASA utilized proportional counters for measuring the muon density through two acquisition modes \cite{AGASA}. The first method, known as the on-off density method, determines the density by counting the number of hit counters based on the assumption that the number of particles incident on each counter follows a Poisson distribution around the mean value. This approach is comparable to the binary mode of the UMDs located at the Pierre Auger Observatory. The second method, called the analogue density method, calculates the density by dividing the total energy losses in all counters by the average energy loss of a vertically traversing muon. This technique is similar to the ADC method mentioned earlier in the UMDs located at the Pierre Auger Observatory. We illustrate the reconstruction methods proposed based on the UMDs of the Pierre Auger Observatory, but note that these methods can be extended to any muon detector with similar configurations.
	\paragraph{} The size of an air shower in surface arrays is characterized by fitting the MLDF. This can be achieved by maximizing a likelihood function and evaluating it at a reference distance \cite{Daniel:08}. The likelihood used is specific to each detector and is based on the response of the detector to incoming particles. For the UMDs at the Auger Observatory, a global estimator of the shower muon density is obtained by reconstructing the MLDF and evaluating it at 450 m. This reference distance of 450 m was proved to be the optimal distance for an array of detectors spaced 750 m apart. This means that muon density fluctuates the least at 450 m, compared to other reference distances, for different samples of the same simulated muon LDF \cite{Diego:16}. In the past, reconstruction methods depended only on information from the binary mode. The time-independent likelihood method is the most straightforward method to reconstruct the MLDF with binary mode data \cite{Daniel:08}. Stations are divided into three different classes based on the number of activated segments, and an approximate likelihood is used for each class. This single-window likelihood method can reconstruct the MLDF with an array of segmented detectors without considering detector timing, using the exact likelihood of the number of muons in a detector given the number of segments with a signal \cite{Daniel:15}. The time-dependent reconstruction method, on the other hand, uses detector timing and a profiling technique to reconstruct the MLDF from binary mode data \cite{Diego:16}. This paper presents a hybrid reconstruction method suitable for a detector with both a binary mode and an ADC mode. Additionally, it proposes a new reconstruction method that uses only the ADC information.
	\paragraph{} Section \ref{sec:2} describes the characterization of the output of the ADC. In Section \ref{sec:3}, the simulation of the Monte Carlo shower library and the ADC mode is discussed. Section \ref{sec:4} outlines the likelihoods that were utilized for the reconstruction process. The performance of the two newly introduced methods is discussed in Section \ref{sec:5}, along with a comparison with the previous reconstruction methods. The paper concludes in Section \ref{sec:6}.
\section{Characterization of the ADC output} \label{sec:2}
\begin{figure}[h!]
    \centering
        \includegraphics[width=0.5\textwidth]{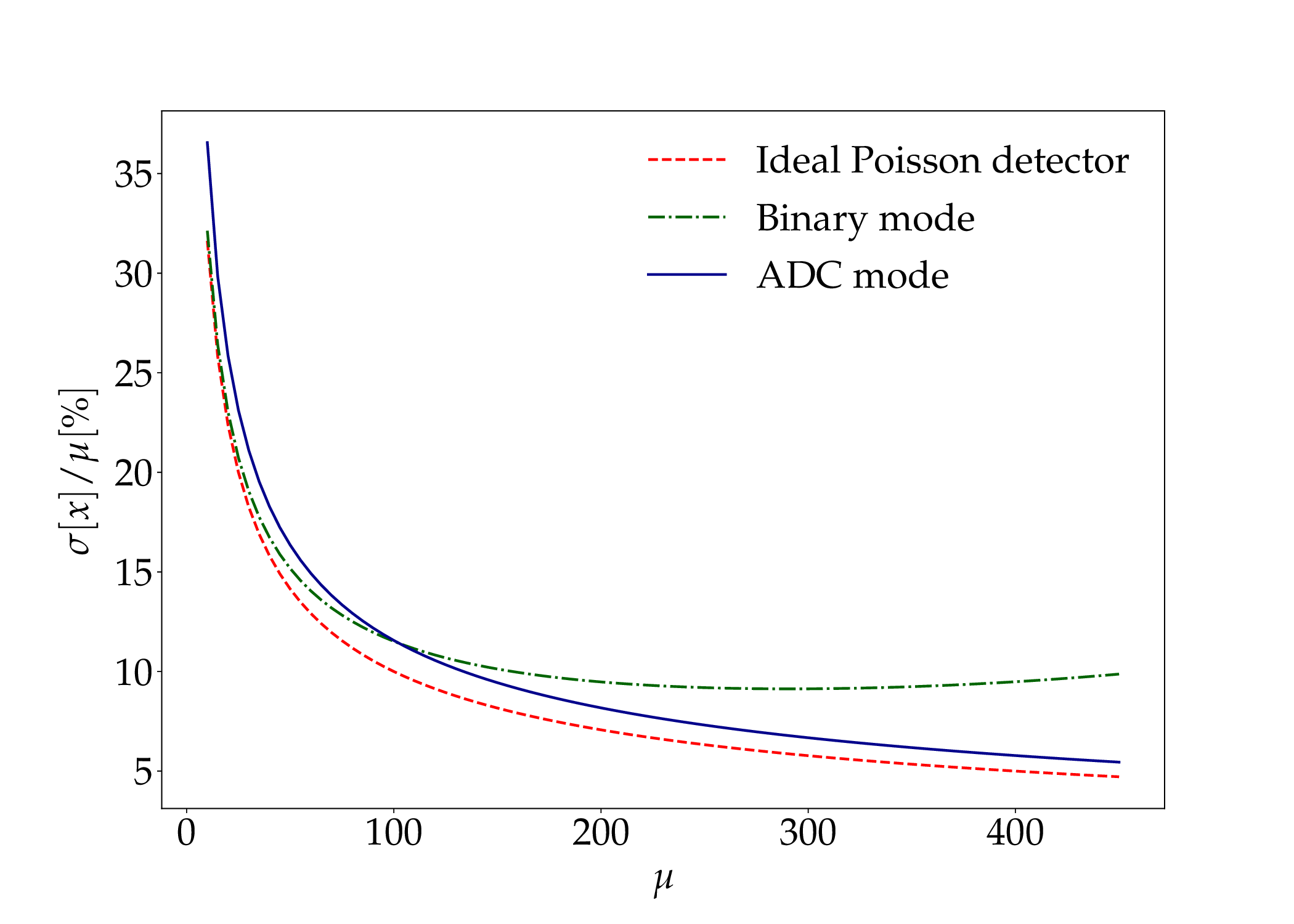}
        \caption{Comparison of the detector resolution for the ADC mode, binary mode and an ideal Poisson detector. The variable $x$ in $\sigma[x]$ refers to the estimator of the number of muons corresponding to each mode.}
        \label{det_res}
    \end{figure}

    \paragraph{} The ADC output is linear and can be expressed as the arithmetic sum of individual signals \cite{Eng:20}. The overall charge associated with $n$ muons can be represented as,
    \begin{equation}\label{Charge}
        Q=\sum\limits_{i=1}^{n}q_i,
    \end{equation}
    \noindent where $q_i$ represent the charge of a single muon. The number of muons that hit the detector ($n$) follows a Poisson distribution ($P(n|\mu)$) with parameter $\mu$.
    \begin{equation}
        P(n|\mu)=\exp{(-\mu)}\dfrac{\mu^n}{n!}.\\ \label{Poisson}
    \end{equation}
    \paragraph{} The average number of muons in equation \eqref{Poisson} is given by $\mu=\rho_\mu A\cos\theta$, where $\rho_\mu$ is the mean muon density of an air shower on the shower plane, $A$ the active area of the UMD module, and $\theta$ the zenith angle of the air shower. The probability density function that characterizes the total output of the detector can be expressed as,
    \begin{equation}\label{totalsig}
        f(Q|\mu) = \sum_{n=0}^{\infty} f(Q|n)\times P(n|\mu),
    \end{equation}
    \noindent where $f(Q|n)$ corresponds to the total charge distribution of $n$ muons. The mean and standard deviation of $Q$ are,
    \begin{eqnarray}
        \langle Q \rangle &=& \mu \langle q \rangle, \label{Mean1}\\
        \sigma^2[Q] &=& \mu \left (\varepsilon^2[q]+1 \right ) \langle q \rangle^2.\label{SD1} 
    \end{eqnarray} 
    \paragraph{} Here $\langle q \rangle$ is the mean charge corresponding to a single incident muon and $\varepsilon[q]$ corresponds to the relative error of the charge deposited by a muon, which is the ratio of the standard deviation to the mean ($\varepsilon[q]=\sigma[q] / \langle q \rangle$). The estimator of the number of muons $\widehat{\mu} = Q / {\langle q \rangle} $ is an unbiased variable with variance $\sigma^2[\widehat{\mu}] = \sigma^2[Q]/\langle q \rangle^2$.
    \paragraph{} As stated in Ref.~\cite{Brian:15}, the uncertainties in estimating the muon density of air showers are dominated by Poisson fluctuations in the muon content. Notably, equation \eqref{SD1} highlights that the ADC contributes to the uncertainty of an ideal Poissonian detector, as the fluctuations in $\widehat{\mu}$ are greater than those expected from the Poisson distribution. As mentioned in the introduction, the resolution of the binary mode plateaus as more particles hit the detector, while the resolution of the ADC is better at higher particle densities. To provide a clearer explanation of the detector's resolution, we compare the resolution of the ADC mode ($\sigma[\widehat{\mu}]/\mu$) with the binary mode and an ideal Poisson detector in Figure \ref{det_res}. In order to calculate the resolution of the binary mode, we utilize the single window (also known as the infinite window strategy), which was developed in Ref.~\cite{Daniel:15}. As depicted in the plot, the binary mode demonstrates similar performance to that of an ideal Poisson detector at low numbers of muons. However, as the number of muons increases, the ADC mode begins to outperform the binary mode.
    \paragraph{} For the ADC mode of the UMDs at the Pierre Auger Observatory we consider the charge of a single muon to be a random variable whose logarithm is normally distributed and the probability density function of charge can be represented using a 2-parameter log-normal distribution of the form,
    \begin{equation}\label{lognormal}
        f(q) = \textrm{LN}(m,\theta^2) \equiv \dfrac{1}{\sqrt{2\pi}\, \theta \, q}\exp\left[-\dfrac{(\ln q-m)^2}{2\theta^2}\right], 
    \end{equation}
    \noindent where $m$ is the scale parameter and $\theta$ is the shape parameter, and can be calculated in terms of $\langle q \rangle$ and $\varepsilon[q]$ as,
    \begin{eqnarray}
        m &=& \ln \left[ \dfrac{\langle q \rangle}{\sqrt{1+\varepsilon^2[q]}} \right],\\
        \theta &=& \sqrt{\ln(1+\varepsilon^2[q])}.
    \end{eqnarray}
    \paragraph{} If the shape parameter remains small ($\theta^2 \lesssim 1$), which is the case for the UMDs at the Pierre Auger Observatory, the sum of $n$ log-normal terms is expected to behave like a log-normal variable, regardless of the value of $n$ as cited in Ref.~\cite{Romeo:03}. When at least one muon is detected ($n \geq 1$), the distribution $f(Q)$ for a total charge $Q$ corresponding to $n$ muons is obtained as,
    \begin{equation}\label{Logsum}
        f(Q|n)\cong \textrm{LN}(m_n,\theta_n^2).
    \end{equation}
    \noindent The log-normal character of $Q$ is verified using simulations (see \ref{App1}).
    \paragraph{} The parameters $m_n$ and $\theta_n$ are estimated from the parameters corresponding to the distribution function of one incident muon ($m$ and $\theta$).
    \begin{eqnarray}
        m_n &=& m+\dfrac{\theta^2}{2}+\ln\left[\dfrac{n}{\sqrt{1+\dfrac{\exp(\theta^2)-1}{n}}}\right]\label{mn},\\
        \theta_n &=& \sqrt{\ln\left[1+\dfrac{\exp(\theta^2)-1}{n}\right]}. \label{thetan}
    \end{eqnarray}
    \paragraph{} In the linear region of the ADC channel, the probability density function of $Q$ can be derived from equations \eqref{totalsig} and \eqref{Logsum}  and can be expressed as a compound distribution of the form,
    \begin{eqnarray}\label{convdist}
        \begin{split}
            f_\textrm{C}(Q|\mu) &\cong \delta(Q) \exp(-\mu) + \sum_{n=1}^{\infty} \dfrac{1}{\sqrt{2\pi}\, \theta_n \, Q }\, \exp{(-\mu)}\dfrac{ \mu^n}{n!}\\
            &\quad \times \exp\left[-\dfrac{\left(\ln Q-m_n\right)^2}{2\theta_n^2}\right]\Theta(Q), 
        \end{split}
    \end{eqnarray}
    \noindent \sloppy where $\delta(Q)$ represents the Dirac delta function, accounting for the possibility of zero signal due to no muons and $\Theta(Q)$ is the Heaviside function. In the worst-case scenario where all muons arrive in the same time bin, a single 10 m$^2$ module causes the LG channel to saturate at about 362 muons, while the HG channel saturates at a much lower value of around 85 muons \cite{Eng:21}. As the UMD consists of three such modules, the detection range is even wider.
    \paragraph{}In regions of higher particle density, i.e., for large values of $\mu$ (or $\widehat{\mu}$), it is expected that the probability density function of total charge for a given value of $\mu$ can be approximated by a Gaussian distribution (refer to \ref{App2}).
    \begin{equation}
        f_\textrm{G}(Q|\mu) \cong \dfrac{1}{\sqrt{2\pi}\, \sigma[Q]}\exp\left[-\dfrac{(Q-\langle Q \rangle)^2}{2\, \sigma^2[Q]}\right], \label{Gausseq}
    \end{equation}
    \noindent where $\langle Q \rangle$ and $\sigma[Q]$ can be obtained from equations \eqref{Mean1} and \eqref{SD1} respectively. 
\section{Simulations}\label{sec:3}
\begin{figure}[h!]
        \centering
        \includegraphics[width=0.5\textwidth]{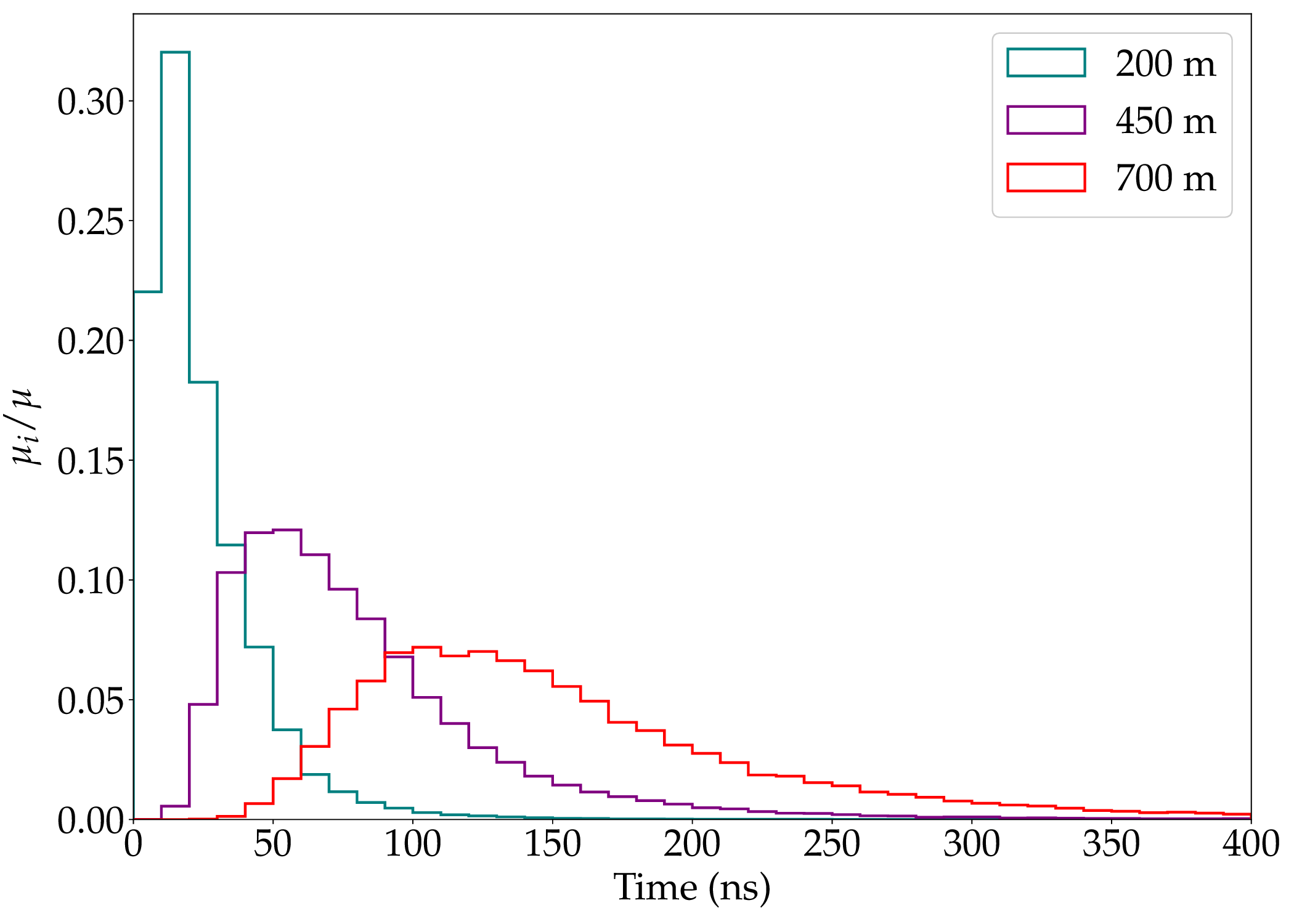}
    	\caption{Average muon arrival time histogram showing the fraction of muons in each time bin at three different distances to the shower axis for iron showers at a zenith angle of 30$\degree$ and energy of $10^{18.5}$ eV.}\label{TD}
    \end{figure}

    \paragraph{} To generate the Monte Carlo shower library for developing the reconstruction method, we used CORSIKA v7.7100 \cite{Heck:97} with the high-energy hadronic interaction model EPOS-LHC \cite{Tanguy:1} and the low-energy hadronic interaction model FLUKA \cite{Fluka:1, Fluka:2}. We simulated showers of iron and proton primaries with energies in the range of $\log_{10}(E/\textrm{eV}) = [17.5,19]$ in steps of $\log_{10}(E/\textrm{eV}) = 0.25$ for zenith angles of $30\degree$ and $45\degree$. The azimuth angles were randomly generated within the range of $-180\degree$ to $180\degree$. To reduce the number of tracked particles, a statistical thinning of $10^{-6}$ was applied during simulation. For each energy and zenith angle, we produced 35 proton and 30 iron showers. Although there is experimental evidence of a muon deficit in simulated air showers \cite{Deficit}, this does not affect the comparison of different strategies to reconstruct the MLDF. This deficit in simulations seems to be related to the lack of knowledge of hadronic interactions at higher energies, as the high-energy hadronic interaction models used in simulations extrapolate the accelerator data to cosmic ray energies.
    \paragraph{} From simulated showers, we obtained the average Muon Lateral Distribution Function (MLDF) and the average arrival time distribution of muons as a function of their distance to the shower axis, for each primary energy, zenith angle, and primary. To account for soil shielding effects, we selected muons with energies greater than $1$ GeV$/{\cos\theta_\mu}$. We fitted the average MLDFs with a KASCADE-Grande MLDF \cite{Apel:10} to obtain an input MLDF. In Figure \ref{TD}, we present the arrival time histogram for iron showers at a zenith angle of 30$\degree$ and energy $10^{18.5}$ eV at three different distances. These histograms show the fraction of particles arriving in 10 ns time bins with respect to the total number of muons. We observed that muons arrive more evenly spaced in time farther away from the shower axis.
    \paragraph{} The MLDF fit and muon time distribution averages obtained for each primary particle, energy level, and zenith angle serve as input for the detector simulations. The binary mode was simulated as described in Ref.~\cite{Daniel:15} and includes the pile-up effect simulation. The muons are counted within a 40 ns time window.
    \paragraph{} The simulation program, which was originally developed in Ref.~\cite{Daniel:15}, has now been expanded to include the simulation of the ADC. The normalized signal as a function of time generated by a single muon can be characterized as a function resembling a log-normal distribution, denoted as $s(t,t_0)$.
    \begin{equation}\label{signal}
    \begin{split}
    s(t,t_0) &= \frac{\Theta(t-t_0)}{\sum\limits_{i=1}^{N} \exp\left[-\frac{\ln^2((t_i-t_0)/\tau)}{2\theta_\textrm{T}^2}\right] \frac{\Delta t}{t_i-t_0} \, (t-t_0)}\\
    &\quad \times \exp\left[ {-\frac{\ln^2((t-t_0)/\tau)}{2\theta_\textrm{T}^2}}\right].
    \end{split}   	
    \end{equation}
\noindent where $\Theta(t)$ is the Heaviside function, $\Delta t$ is the sampling time (6.25 ns for this work), $t_0$ is the signal start time, $m_\textrm{T} = \ln(\tau/\textrm{ns})$ and $\theta_\textrm{T}$ are the parameters of a log-normal-like function, and $N$ is the number of samples of the signal. Note that $s(t,t_0)$ has the dimension of [time]$^{-1}$.
    \paragraph{} Once the ADC reaches saturation, the linear proportionality between the signal charge and the number of injected muons is lost. To determine the saturation level of the ADC signal, one can calculate the maximum of the function $s(t,t_0)$ for a specific number of muons required to saturate the ADC channel ($N_\mu$).
    \begin{equation}\label{Satpul}
	   S_\textrm{L} =  \frac{N_\mu \langle q \rangle}{\sum\limits_{i=1}^{N} \exp\left[{-\frac{(\ln^2((t_i-t_0)/\tau)}{2\theta_\textrm{T}^2}}\right] \frac{\Delta t}{t_i-t_0}} \dfrac{\exp\left ({\dfrac{\theta_\textrm{T}^2}{2}}\right )}{\tau}.
    \end{equation}
    \paragraph{} In this study, we focus solely on the LG channel of the UMD at the Pierre Auger Observatory. The saturation level is determined by the value of $N_\mu=1086$, which represents three times the number of muons required by a single module to reach saturation \cite{Eng:21}. The parameters of the log-normal-like function are selected as $\ln(\tau/\textrm{ns})=4$ and $\theta_\textrm{T}=0.4$. These values correspond to a particular set of laboratory measurements performed using four 4-m-long scintillator strips inside an aluminum container. Optical fibers were glued to each strip and coupled through an optical connector to a calibrated SiPM array, which consisted of 64 SiPMs connected to the front end of the electronics kit and sealed into a PVC box. A muon telescope was used to trigger the electronics when the muon crossed the four bars. Muon pulses measured with the ADC were recorded at different positions of the muon telescope along the bars, and the average signal as a function of time of one muon was obtained (see reference \cite{Botti:19} for more details). Note that any other suitable parameter values for the  of the log-normal-like function would produce similar outcomes. Detector saturation establishes the upper limit of the ADC detection range.

\begin{figure}[h!]
        \centering
    	\includegraphics[width=0.5\textwidth]{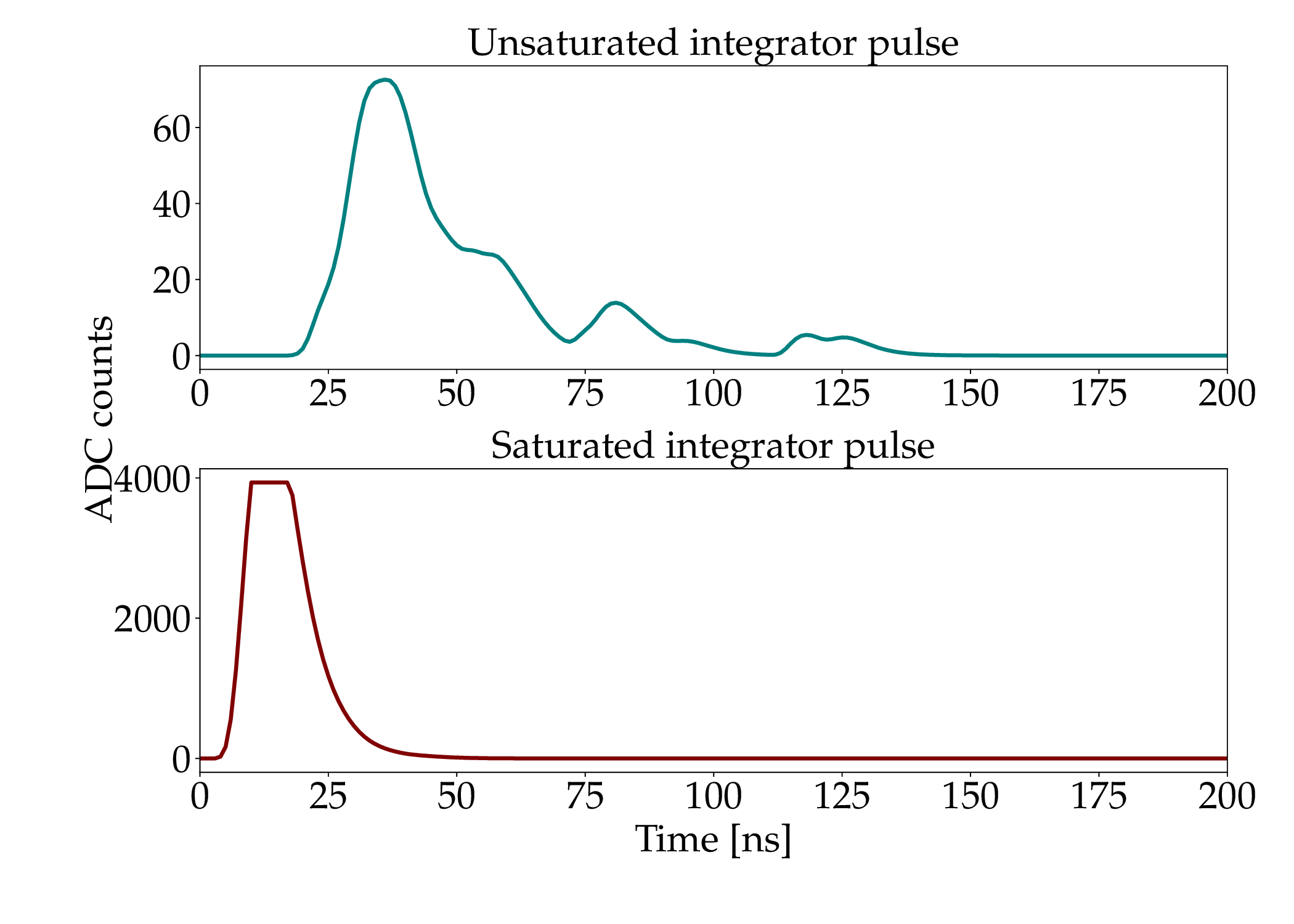}
	    \caption{Simulated unsaturated and saturated pulses at the LG channel of the ADC for an iron shower of $10^{19}$ eV and $\theta = 30^\circ$. The distances to the shower axis of the unsaturated and saturated pulses are $\sim 876$ m and $\sim 211$ m, respectively.}\label{Pulses}
    \end{figure}

    \paragraph{}The ADC signal in units of ADC counts/time corresponding to $n$ muons is obtained by considering the unsaturated signal,
    \begin{equation}
        S_\textrm{US}(t)=\sum_{i=1}^{n} q_i\, s(t,t_i),
    \end{equation}
    \noindent where the $n$, $t_i$, and $q_i$ values are obtained by sampling the Poisson distribution, the corresponding average muon arrival time distribution, and the charge distribution of one muon (see equation (\ref{lognormal})), respectively. Finally, the signal is obtained from the following expression,
    \begin{equation}
        S(t)=\left\{ 
        \begin{array}{ll}
           S_\textrm{US}(t)  &  \ \ \ \textrm{if} \ \ \ S_\textrm{US}(t)<S_\textrm{L} \\[0.4cm]
           S_\textrm{L}  & \ \ \ \textrm{if} \ \ \ S_\textrm{US}(t) \geq S_\textrm{L}
        \end{array}
        \right..
    \end{equation}
    \paragraph{}The values of the parameters corresponding to the single muon charge distribution of equation (\ref{lognormal}) considered are $m = 5$ and $\theta = 0.5$ \cite{Botti:19}. In this case as well, these values correspond to the same particular set of laboratory measurements representing the given muon detector. Figure \ref{Pulses} shows an unsaturated and saturated pulse for the aforementioned scenario.
    \paragraph{}The total signal or charge in a given detector ($Q$) in equation (\ref{Charge}) is obtained by adding the signal evaluated in each discrete time value, obtained according to the sampling time of the ADC, multiplied by $\Delta t$. 
\section{Likelihoods and the reconstruction method}\label{sec:4}
\begin{figure}[h!]	
    \centering
        \includegraphics[width=0.5\textwidth]{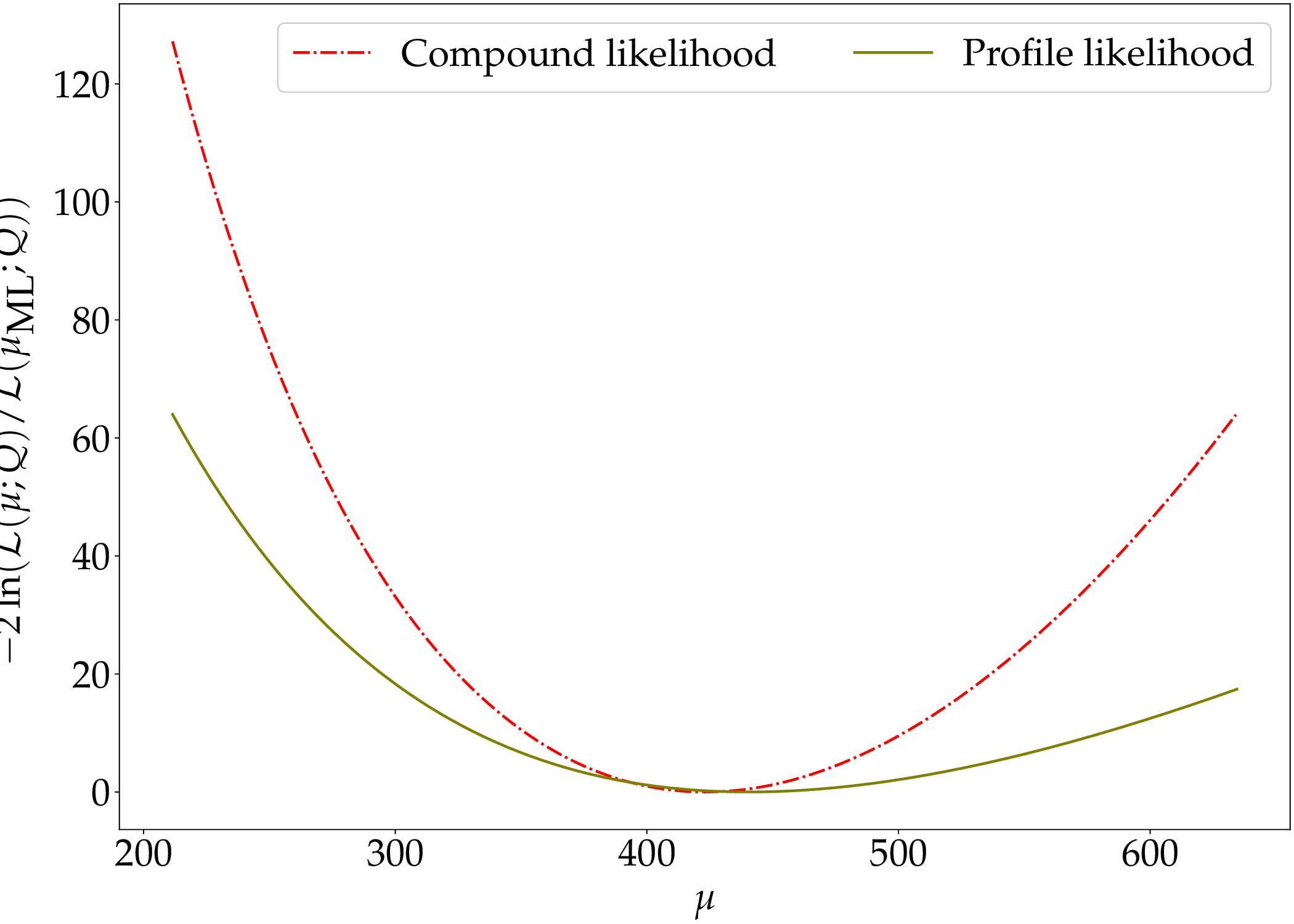}
        \caption{Comparison of different likelihoods used in this work. The plots correspond to iron primary at 10$^{18}$ eV at 30$\degree$ zenith angle. The maximum likelihood estimator of $\mu$ is obtained from Profile likelihood and has the value $\widehat{\mu} = 423$ }
        \label{likeplots}
    \end{figure}

    \paragraph{} This paper examines two reconstruction methods: the ADC reconstruction method, which relies on information from the ADC only, and the ADCProfile reconstruction method, which combines ADC and binary outputs. The ADCProfile reconstruction method is an extension of the Profile reconstruction method that was developed in \cite{Diego:16}.
    \paragraph{} The MLDF, denoted by $\mu(r;\Vec{p})$, is dependent on two free parameters in $\Vec{p}=(\mu_0,\beta)$ and the distance between the detector and the shower axis $(r)$. Here, $\mu_0$ represents the normalization parameter, while $\beta$ is the slope. The MLDF is factorized by $\mu_0$ and a form function $g(r;\beta)$ that rapidly decreases with the distance to the shower axis,
    \begin{equation}
        \mu(r;\Vec{p})  = \mu_0 \frac{g(r;\beta)}{g(r_0;\beta)},
    \end{equation}
    \noindent where $g(r;\beta)$ is,
    \begin{equation}
        g(r;\beta) = \left(\frac{r}{r_1} \right)^{-\alpha}\left(1+\frac{r}{r_1} \right)^{-\beta}\left( 1+\left( \frac{r}{10r_1}\right)^2 \right)^{-\gamma}.
    \end{equation}
    \paragraph{}Fixed values are taken for $\alpha=0.75$, $r_1=320$ m, and $r_0$ is a reference distance, which is taken as 450 m in this work. The $\gamma$ parameter is also a fixed value and is obtained from the input MLDF. It is almost constant with energy, zenith angle, and primary type. It describes the behavior of the MLDF at large distances to the shower axis, where the statistics are low and the detector has few muons or is non-triggered. In this work, we fix $\gamma$ at $-4.18$. The free-fit parameters in $\Vec{p}$ are obtained by minimizing the function,
    \begin{equation}
        -2 \ln \mathcal{L}_\textrm{T} = -2 \sum_{i} \ln \left[\mathcal{L}(\mu(r_i,\Vec{p})) \right]. \label{CI}
    \end{equation}
    \noindent The sum runs over the detectors in this context. To choose $\mathcal{L}$ for ADC reconstruction, the estimated number of muons in the LG channel of the ADC is considered. On the other hand, for ADCProfile reconstruction, $\mathcal{L}$ selection depends on the estimated number of muons in the various time windows of the binary traces.
    \paragraph{}To begin with, let us examine the ADC reconstruction. If the estimated number of muons in the ADC channel, denoted as $\widehat{\mu}$, is less than or equal to 3, the station is labeled as non-triggered. This is done to eliminate any accidental triggers that may be caused by random coincidences. In this case the probability is calculated by integrating equation \eqref{convdist} from $-\infty$ to $Q_{\textrm{min}}$,
    \begin{equation}\label{Silent}
        \begin{split}
            \mathcal{L}(\mu;Q) &= \left( 1+ \frac{1}{2} \sum_{n=1}^{\infty} \erfc \left[ \frac{m_n - \ln(Q_{\textrm{min}})}{\sqrt{2}\, \theta_n} \right] \frac{\mu^n}{n!} \right)\\
            &\quad \times \exp(-\mu).
        \end{split}
    \end{equation}
    \noindent Here $Q_{\textrm{min}}=3 \langle q \rangle$ and $\erfc(x)=1-\erf(x)$, where $\erf(x)$ is the error function. 
    \paragraph{} If a station triggers and the estimated number of muons in the LG ADC channel is less than 200, a compound likelihood function is taken into account,
    \begin{equation}\label{compound}
        \begin{split}
            \mathcal{L}(\mu;Q) &=\sum_{n=1}^{\infty} \frac{1}{\sqrt{2\pi}\, \theta_n \, Q} \exp\left[ \frac{-(\ln Q-m_n)^2}{2\theta_n^2} \right] \\
            &\quad \times \frac{\mu^n}{n!} \exp(-\mu).
        \end{split}  
    \end{equation}
    \paragraph{} When the estimated number of muons in the ADC channel is equal to or greater than 200 and the station is not saturated, a Gaussian likelihood is taken into consideration (refer to \ref{App2}). This is done to improve the performance of the reconstruction.
    \begin{equation}\label{Gaus}
        \mathcal{L}(\mu;Q)=\frac{1}{\sqrt{2\pi\sigma^2[Q]}}\exp\left[-\frac{(Q-\langle Q \rangle)^2}{2\sigma^2[Q]} \right],
    \end{equation}
    \noindent where the mean and the variance are given by equations \eqref{Mean1} and \eqref{SD1} respectively.
    \paragraph{} If a particular station is saturated, we obtain the likelihood by integrating equation \eqref{Gaus} from $Q$ to $+\infty$. Here $Q$ refers to the integral of a pulse, where the upper portion has been clipped (refer to the bottom panel of figure \ref{Pulses}),
    \begin{equation}
        \mathcal{L}(\mu;Q) = \frac{1}{2} \left( 1-\erf \left[ \frac{Q-\langle Q \rangle}{\sqrt{2\sigma^2[Q]}} \right] \right)\label{Erf},
    \end{equation}
    \noindent where the mean and the variance are given by equations \eqref{Mean1} and \eqref{SD1} respectively.
    \paragraph{}Now let us consider the ADCProfile reconstruction method. In this case, the ADC likelihood is utilized when the binary mode of a given muon detector produces a number of bars with a signal ($k$) greater than or equal to 124 in at least one inhibition window. If $k<124$ in all time windows of the trace, the Profile likelihood of Ref.~\cite{Diego:16} is used. This choice of $k$ is motivated by the fact that the relative uncertainty in estimating the number of incident muons that fall in the same inhibition windows is less than $\sim 16\%$ for $k\leq 124$ \cite{Daniel:20}. The non-triggered stations are defined and treated as described in Ref.~\cite{Diego:16}.
    \paragraph{} To compare the various likelihoods considered in this study, we plotted the logarithm of the likelihoods, as shown in Figure \ref{likeplots}. We compared the compound likelihood in equation \eqref{compound} to a profile likelihood from Ref.~\cite{Diego:16}, which was utilized in the ADCProfile reconstruction approach. The plots corresponds to a value of $\widehat{\mu}_{\textrm{ML}} = 423$, which is the maximum likelihood estimator of $\mu$ obtained from Profile likelihood. 
\section{Performance of the reconstruction methods}\label{sec:5}
\begin{figure}[h!]	
    \centering
    \includegraphics[width=0.45\textwidth]{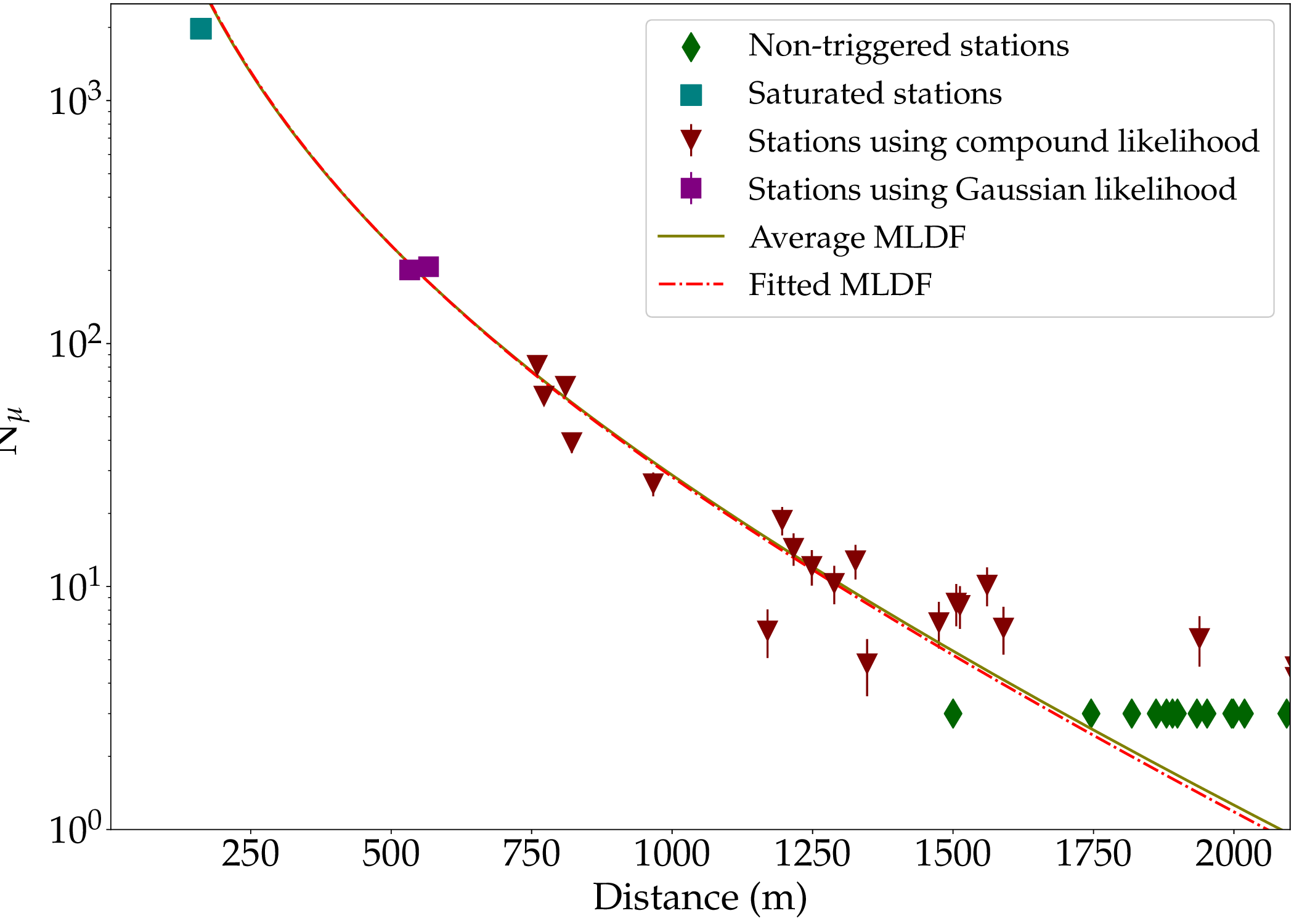}
    \includegraphics[width=0.45\textwidth]{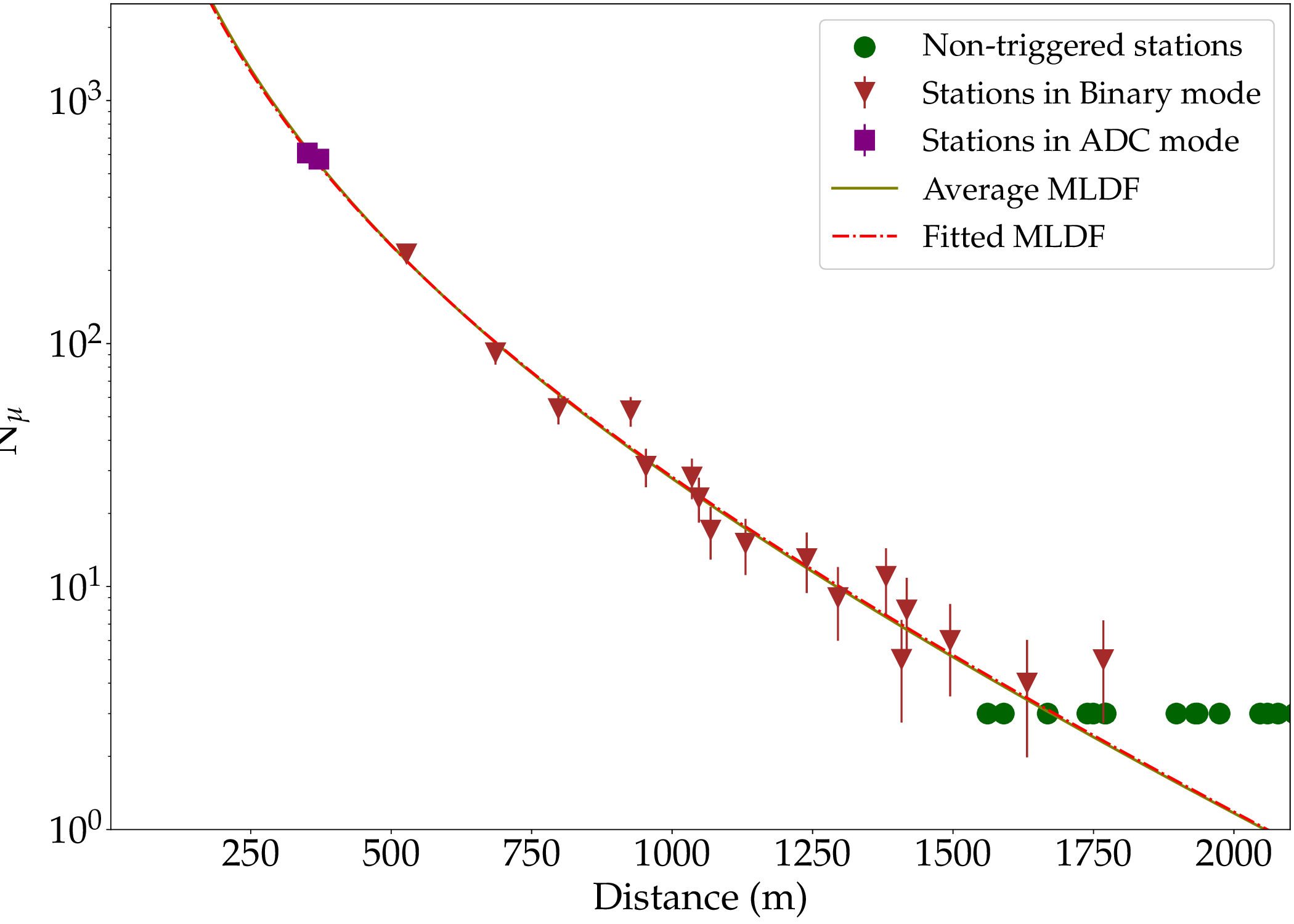}
    \caption{The muon lateral distribution function was fitted to the simulated data using the ADC reconstruction method (top panel) and the ADCProfile reconstruction method (bottom panel). The data correspond to the estimated number of muons in the UMD calculated with a simulation of an iron primary at $10^{19}$ eV and $30\degree$ zenith angle. The solid lines correspond to the MLDF used as the input for the simulation. Note that the plots corresponds to two different events.\label{Eve}} 
    \end{figure}

\begin{figure}[h!]
        \centering
        \includegraphics[width=0.45\textwidth]{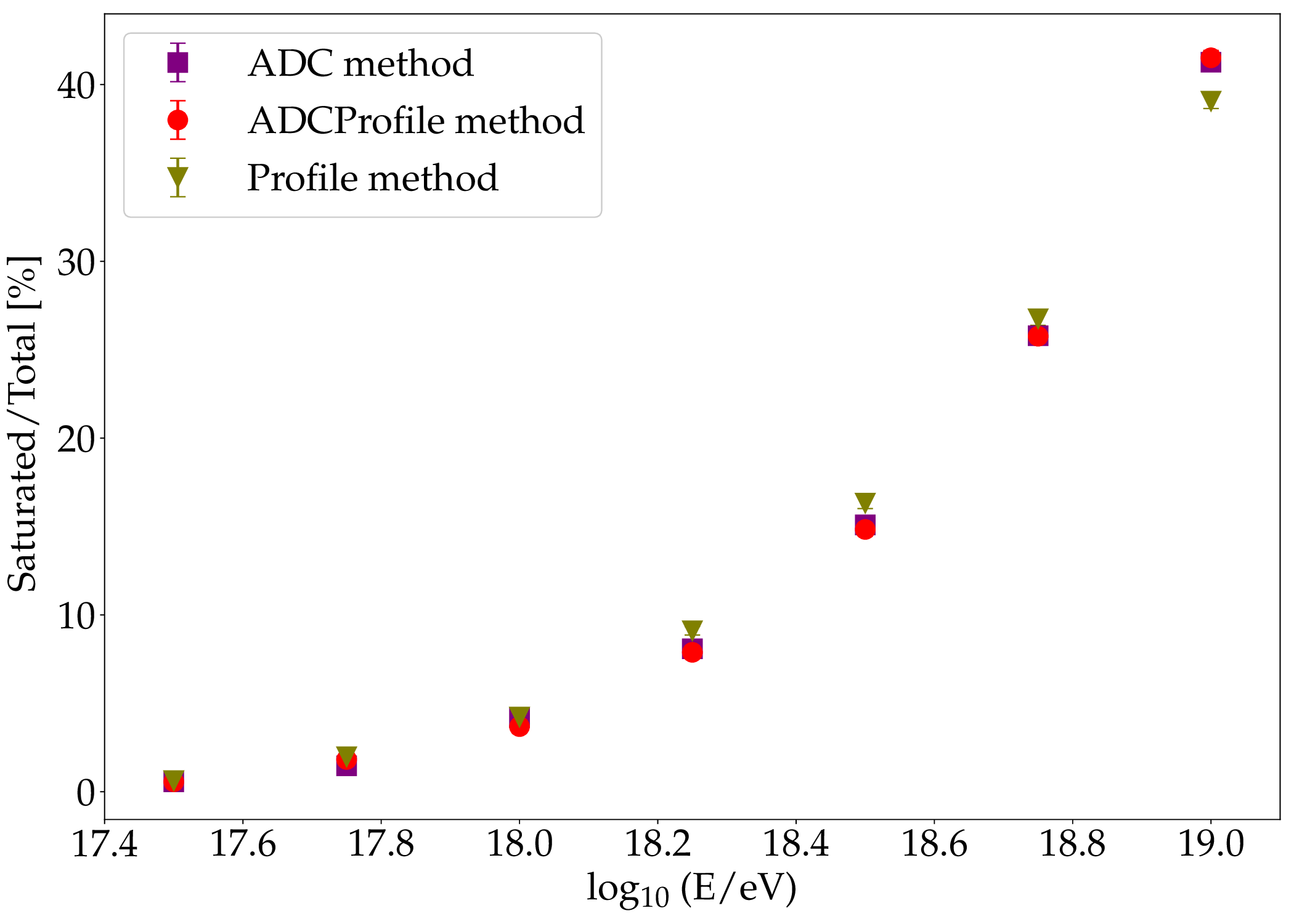}
        \caption{Fraction of saturated events for iron primaries at 30$\degree$ zenith angle as a function of the logarithm of primary energy for ADC, ADCProfile, and Profile reconstruction methods, including the error bars. The error bars are smaller than the marker size.} \label{satfrac}
    \end{figure}

\begin{figure}[h!]
    \centering
    \includegraphics[width=0.45\textwidth]{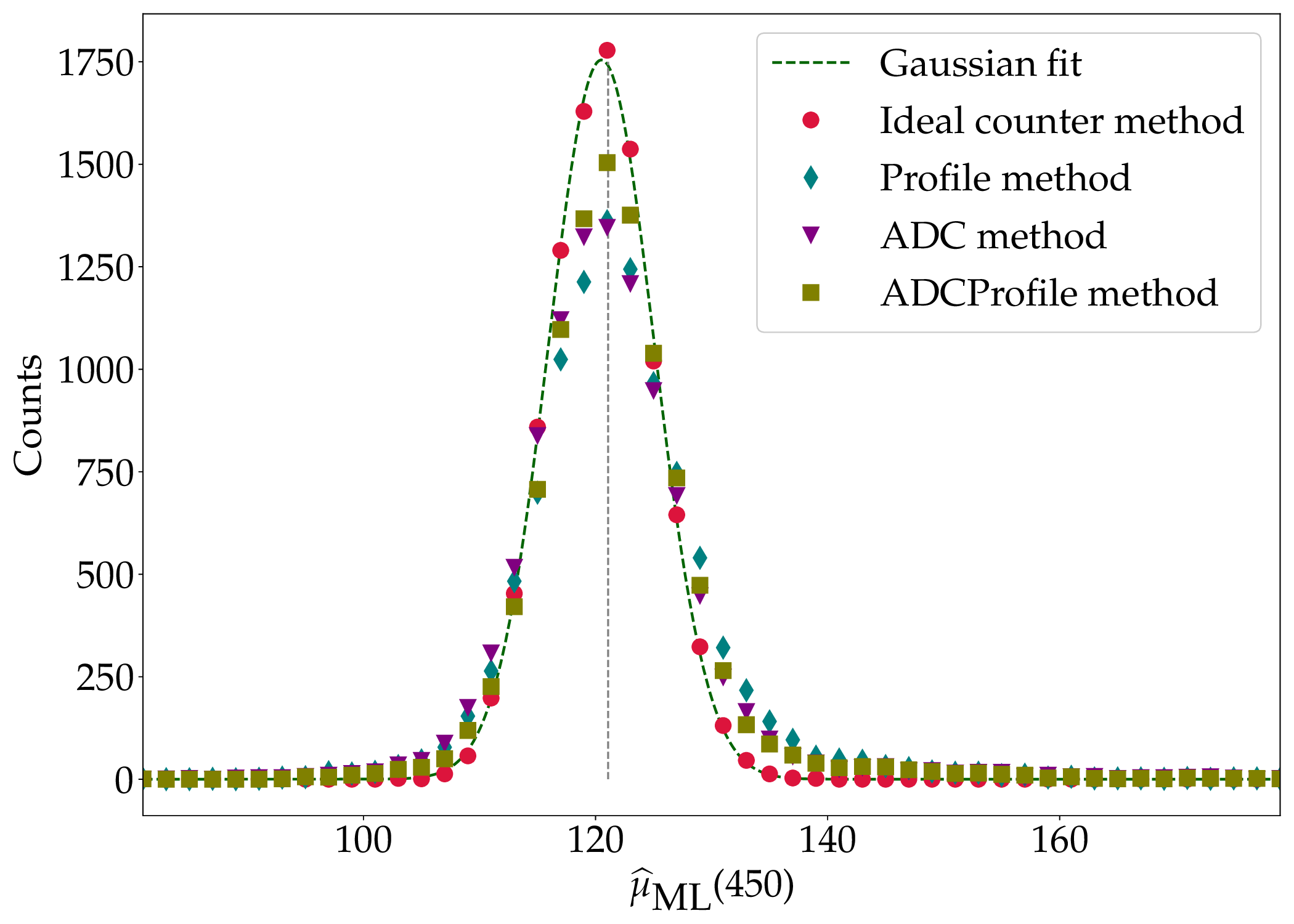}
    \caption{A histogram of the reconstructed $\widehat{\mu}_{\textrm{ML}}(450)$ for iron primaries at $10^{18.5}$ eV and 30$\degree$ zenith angle for the different reconstruction methods considered. The Gaussian distribution is fitted to the ideal counter data.}\label{nmu}
    \end{figure}

\begin{figure}[h!]
    \centering
    \includegraphics[width=0.45\textwidth]{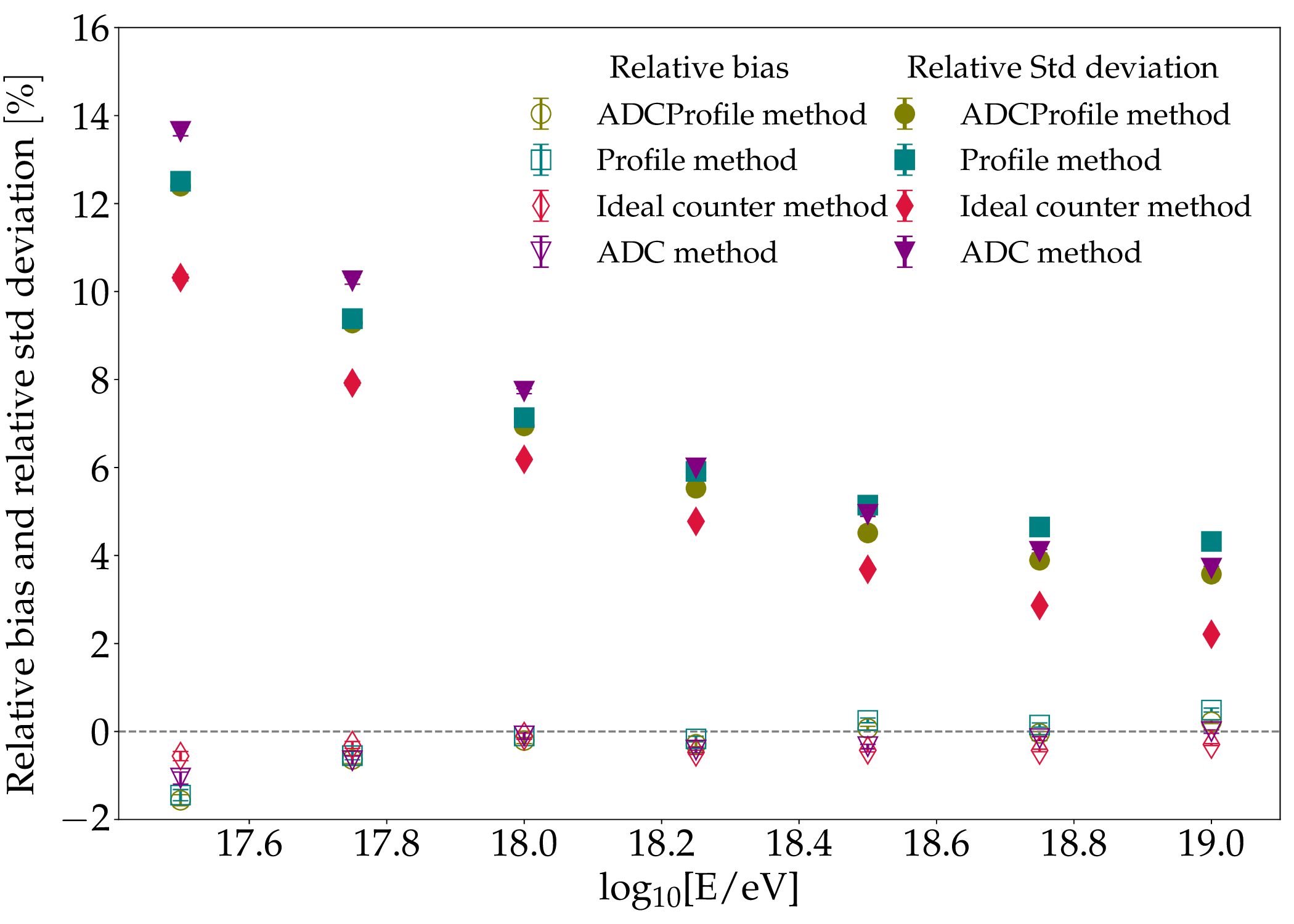}
    \caption{Relative bias and relative standard deviation of the reconstructed $\widehat{\mu}_{\textrm{ML}}$(450) as a function of the logarithm of the primary energy for ADC, ADCProfile, and Profile reconstructions, compared against an ideal counter The plot corresponds to the iron primaries at 30$\degree$ zenith angle. The error bars are smaller than the marker size for all reconstruction methods.} \label{bias}
    \end{figure}

\begin{figure}[h!]
    \centering
    \includegraphics[width=0.45\textwidth]{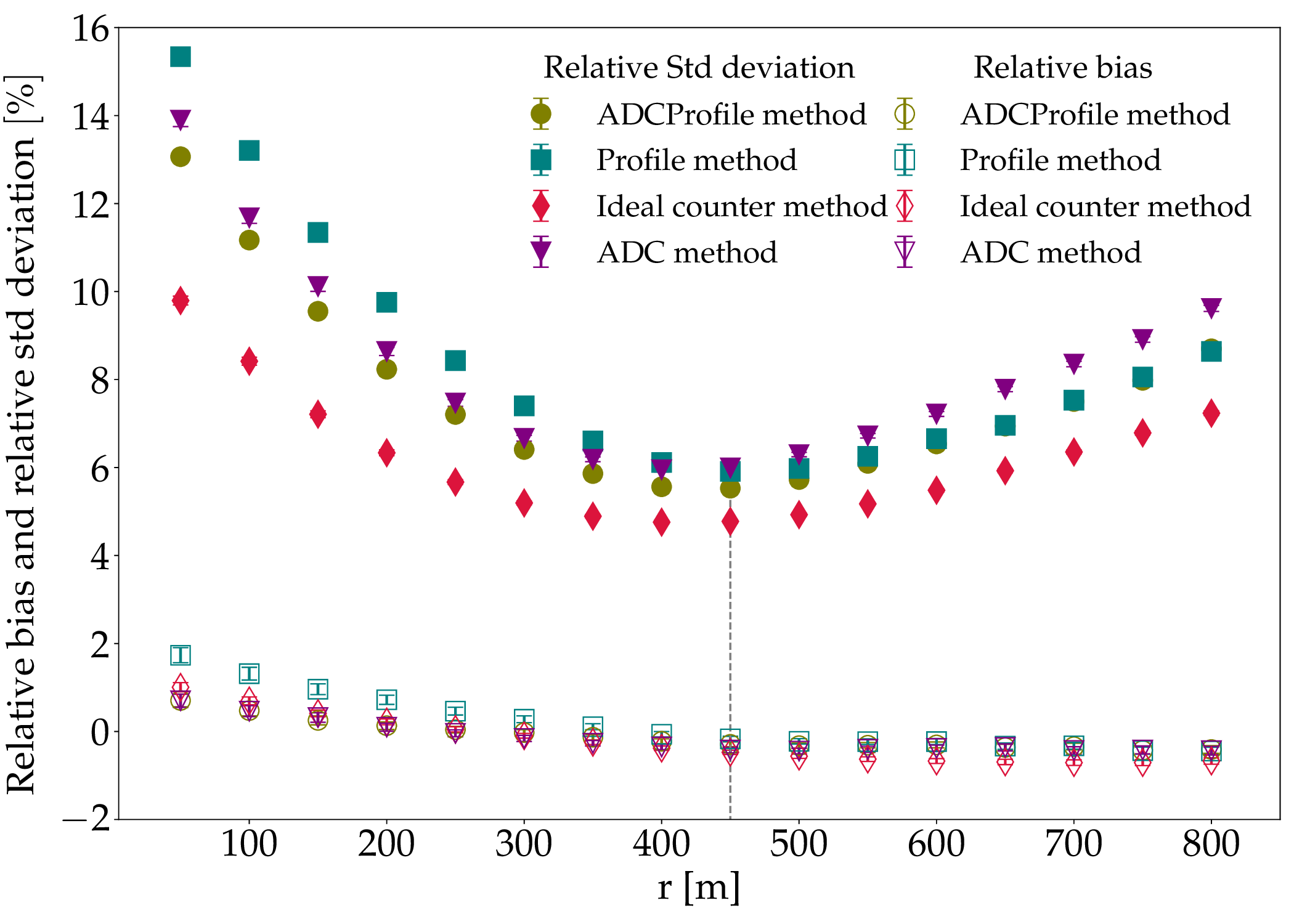}
    \caption{Relative bias and relative standard deviation of the reconstructed $\widehat{\mu}_{\textrm{ML}}$(r) as a function of distance to the shower axis for iron primaries at 30$\degree$ zenith angle and $10^{18.25}$ eV energy The dotted line represents the reference distance of the UMD. The ADC, ADCProfile, and Profile reconstruction methods are compared against an ideal counter. The error bars are smaller than the marker size for all reconstruction methods.} \label{biasvsr} 
    \end{figure}

\begin{figure}[h!]
    \centering
    \includegraphics[width=0.45\textwidth]{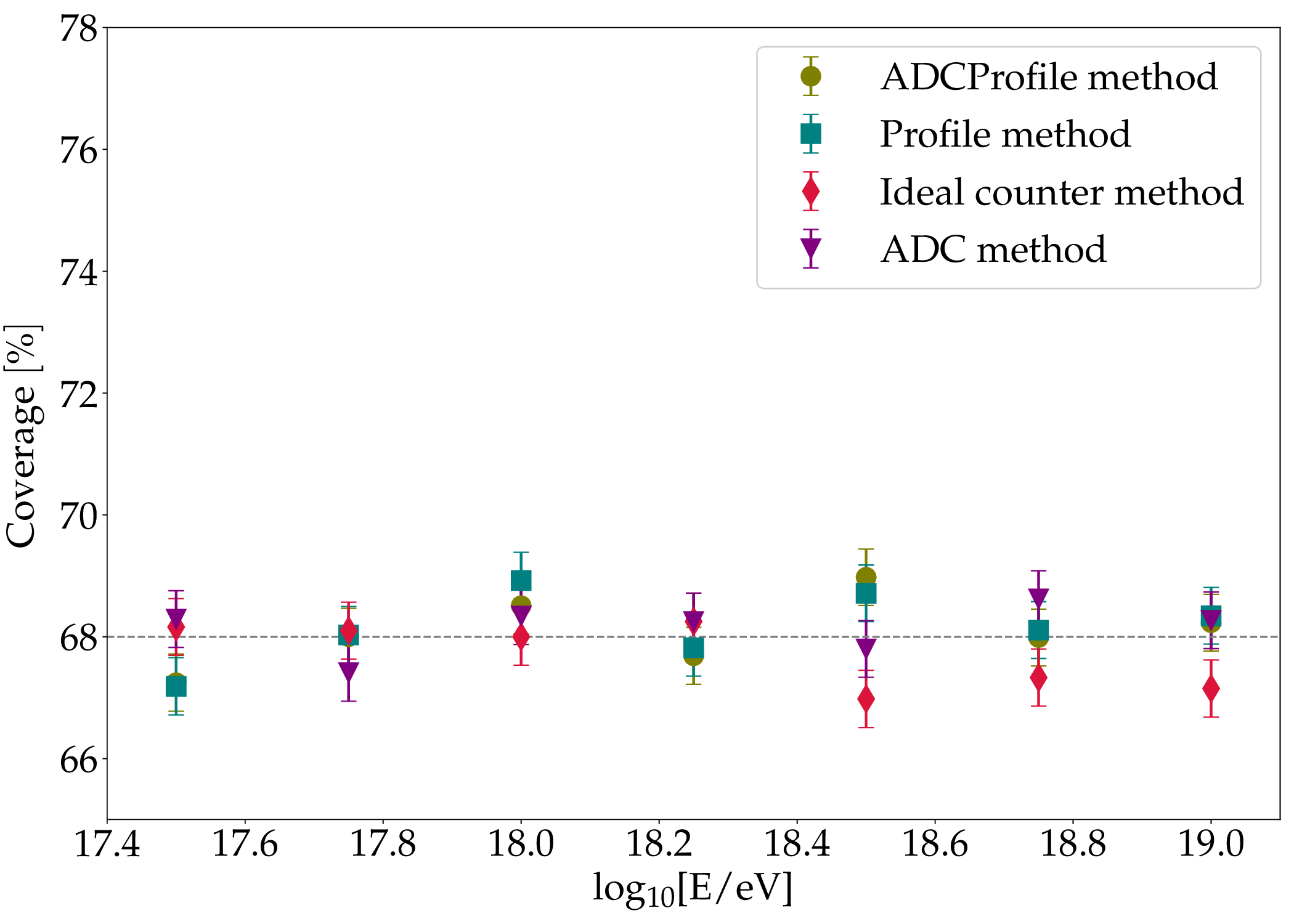}
    \caption{The coverage of the $1\sigma$ confidence interval of $\widehat{\mu}_{\textrm{ML}}$(450) as a function of the logarithm of energy for iron primaries at 30$\degree$ zenith angle. The dotted line at $68\%$ corresponds to the coverage of the $1\sigma$ confidence interval of Gaussian likelihood. The ADC, ADCProfile, and Profile reconstruction methods are compared against an ideal counter.}\label{Cov}
    \end{figure}

\begin{figure}[h!]
        \centering
    		\includegraphics[width=0.45\textwidth]{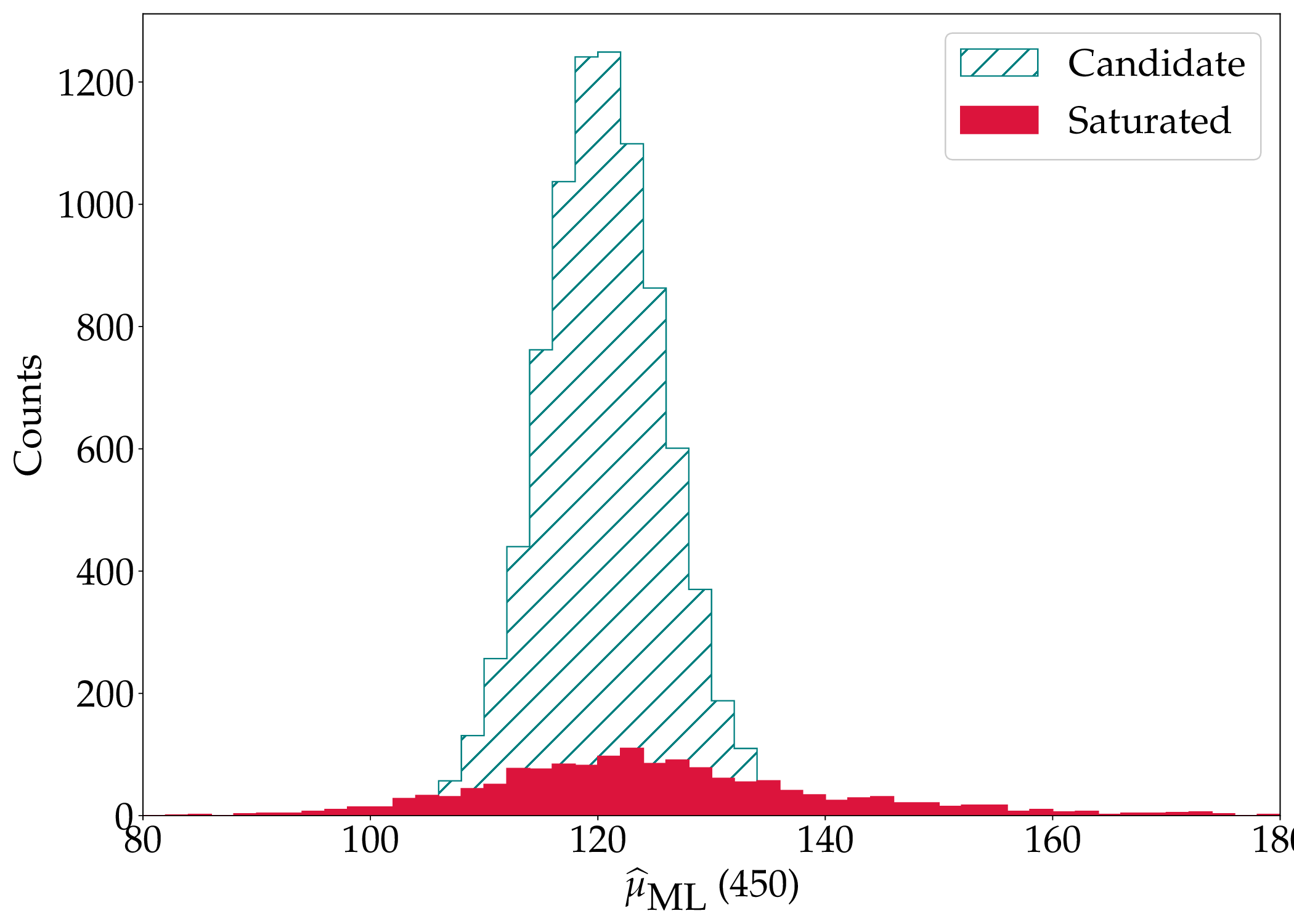}
    		\includegraphics[width=0.45\textwidth]{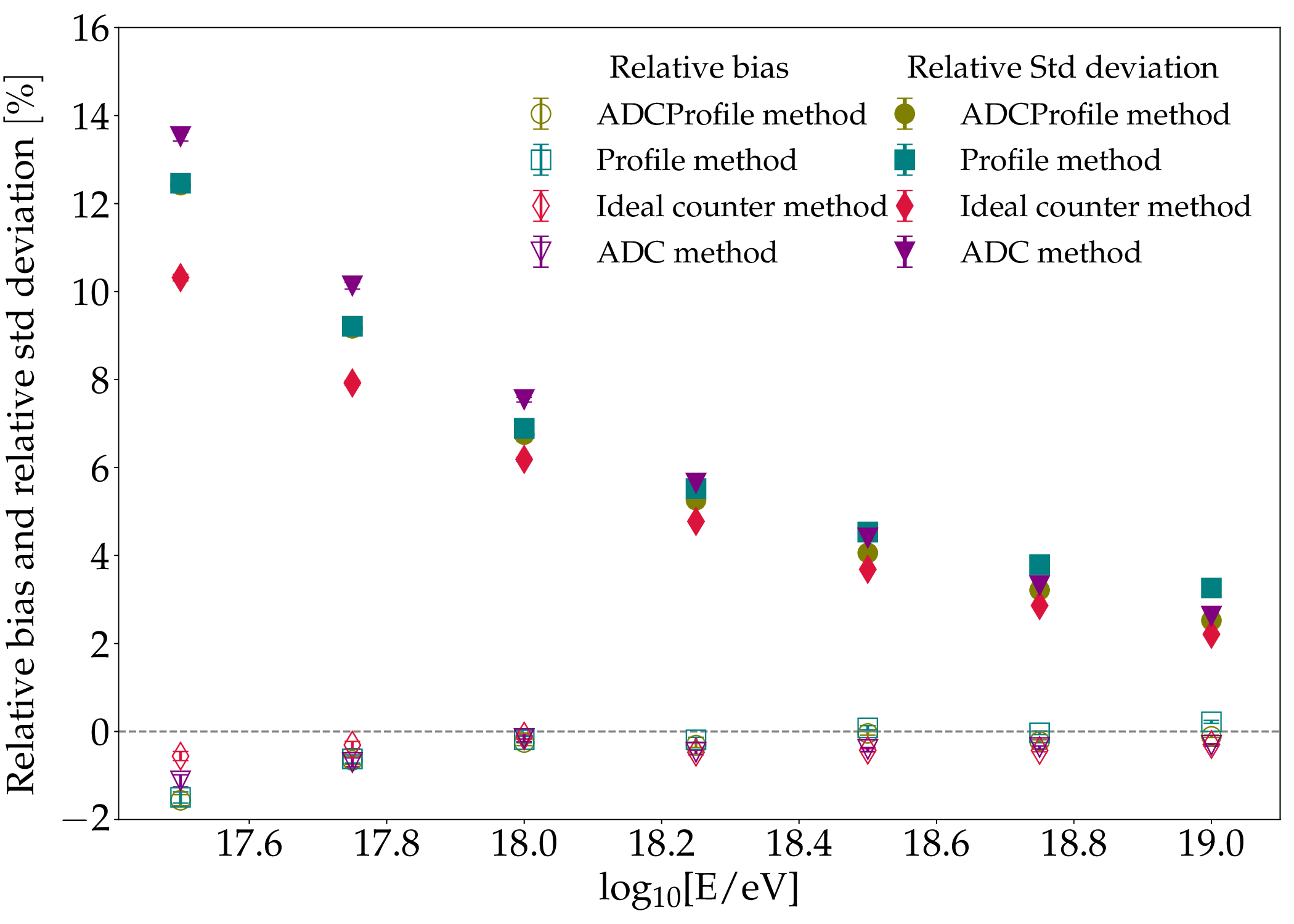}
	    \caption{A comparison of the effects of saturated events in the reconstruction procedure. The top panel shows the histograms of the reconstructed $\widehat{\mu}_{\textrm{ML}}$(450) for iron primaries at $10^{18.5}$ eV energy and 30$\degree$ zenith angle for the ADC reconstruction method. The blue and red histograms corresponds to the reconstructed candidate and saturated events respectively. The bottom panel shows the relative bias and the relative standard deviation of the reconstructed $\widehat{\mu}_{\textrm{ML}}$(450) against the logarithmic energy for iron primaries at 30$\degree$ zenith angle for different reconstruction methods, excluding the saturated events.} \label{Nmu_ADC}
    \end{figure}

\begin{figure}[h!]
        \centering
    	\includegraphics[width=0.45\textwidth]{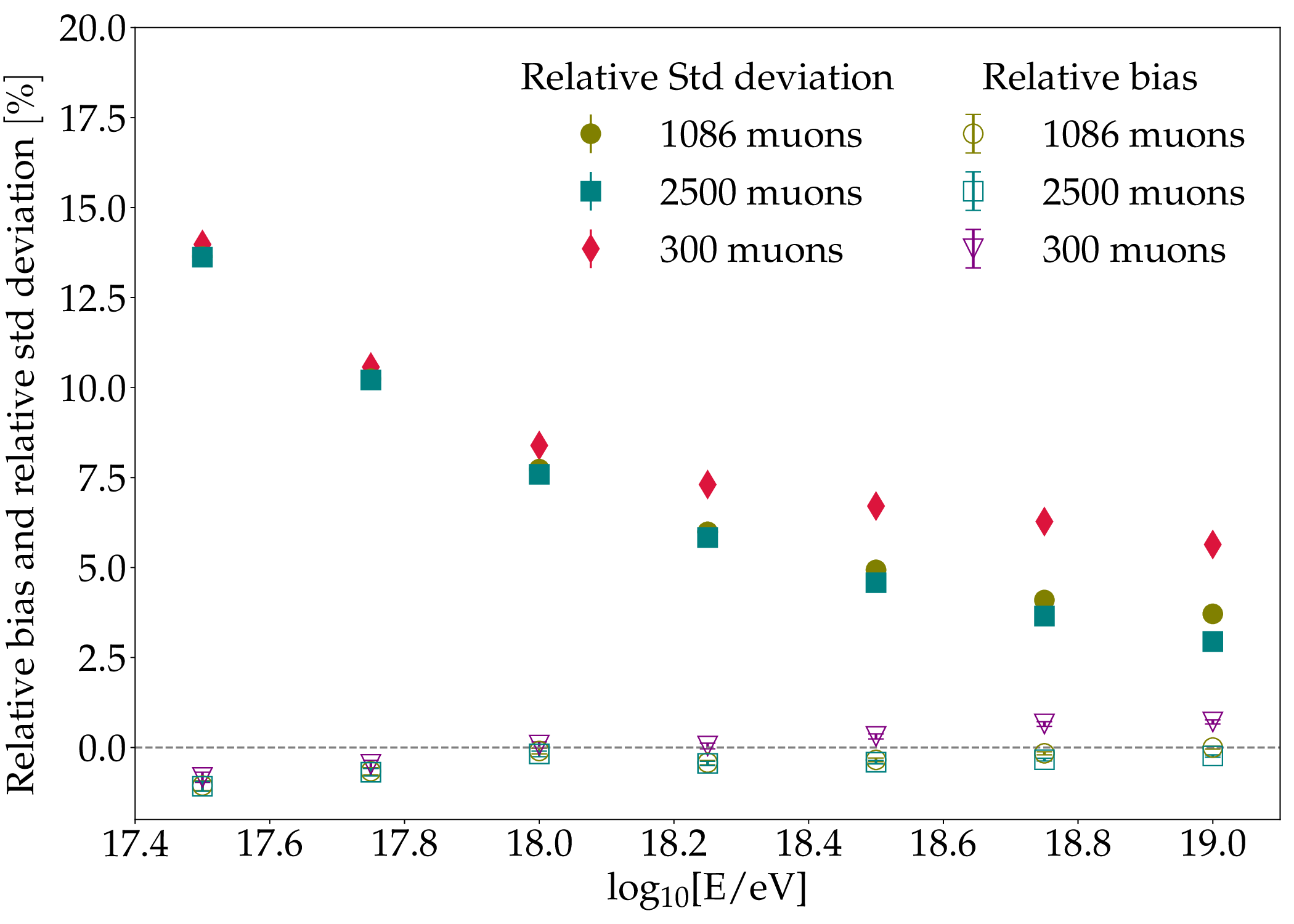}
    	\caption{A comparison of the relative bias and relative standard deviation of the reconstructed $\widehat{\mu}_{\textrm{ML}}$(450) with the ADC reconstruction method at three different saturation levels of the LG channel as a function of the logarithm of energy.} \label{satlev_comp}
    \end{figure}

    \paragraph{} In this section, we will evaluate the performance of the new reconstructions. As already explained in section \ref{sec:3}, to simulate the number of muons arriving at the detector for each shower, we sampled the Poisson distribution with parameter $\mu$ given by the average MLDF. The time distributions were sampled to simulate the arrival times of the muons. Each of these realizations is referred to as an event. We simulated 10,000 events for each primary type, zenith angle, and primary energy, including the simulation of the binary and ADC acquisition modes. The number of muons as a function of the distance to the shower axis for simulated events was adjusted using the Profile, ADC, and ADCProfile likelihoods, and then compared against an ideal counter \cite{Daniel:15}. An ideal counter simply counts the number of muons crossing it, which sets the best-case scenario for the resolution achievable with a muon detector. To minimize the log-likelihoods, we used the numerical minimization software library MINUIT \cite{Minuit}, which is implemented in the ROOT data analysis framework \cite{Root}.
    \paragraph{} Figure \ref{Eve} displays an example of an MLDF fit achieved using the ADC reconstruction method (top panel) and ADCProfile reconstruction method (bottom panel). The presented data depict the number of muons in the UMD for iron primaries at $10^{19}$ eV and 30$\degree$ zenith angle, with each station labeled according to the likelihood utilized in the reconstruction process. The top panel plot indicates that Gaussian likelihood (see equation (\ref{Gaus})) is used for stations near the shower axis, while Compound likelihood (see equation (\ref{compound})) is used for other stations. The plot also shows a few non-triggered stations and a saturated station. On the other hand, the bottom panel plot demonstrates that the ADC reconstruction method is used for stations near the shower axis, while the Profile reconstruction method is used for other stations.
    \paragraph{} Figure \ref{satfrac} displays the percentage of saturated events, as a function of the logarithmic energy. An event is categorized as saturated when it involves at least one saturated muon detector. In both the ADCProfile and ADC approaches, detectors are viewed as saturated when the estimated count of muons in the ADC channel surpasses the saturation limit of the LG channel, as described by equation \eqref{Satpul}. In the Profile reconstruction technique, a detector is regarded as saturated if, in a single time window, a station contains 192 bars with signal. The graph shows a noticeable increase in the percentage of saturated events in relation to primary energy. Moreover, the fraction of saturated events appears to be similar for all the considered reconstruction methods, primarily due to the use of 40 ns wide inhibition windows in binary mode. However, the fraction of saturated events for Profile reconstruction is smaller when the inhibition window is shorter (refer to \ref{App3}).
     \paragraph{} The reconstructed $\widehat{\mu}_{\textrm{ML}}(450)$ (the maximum likelihood estimator of the average number of muons at 450 m from the shower axis) varied in reconstructions of the same shower due to fluctuations in detector signals. Figure \ref{nmu} shows the distributions of the reconstructed $\widehat{\mu}_{\textrm{ML}}(450)$ for iron primaries at $10^{18.5}$ eV and 30$\degree$ zenith angle, considering various reconstruction methods and includes the saturated events. The data corresponding to the ideal counter were fitted with a Gaussian distribution, and the dotted line represents the input value of $\mu$(450). As anticipated, the narrowest distribution corresponds to the ideal counter.
    \paragraph{} To assess the reconstruction performance, it is necessary to compare the input value $\mu(450)$ to the corresponding value fitted to the simulated data. The relative bias is defined as the difference between the input $\mu(450)$ and the average of $\widehat{\mu}_{\textrm{ML}}$(450) calculated over all reconstructions, i.e., $B=\mu(450)-\langle \widehat{\mu}_{\textrm{ML}}(450)\rangle$, and represents the systematic uncertainty in the estimation of $\mu(450)$. Figure \ref{bias} displays the relative bias and relative standard deviation, $\varepsilon(450)=\sigma[\widehat{\mu}_{\textrm{ML}}](450)/\mu(450)$, of $\widehat{\mu}_{\textrm{ML}}(450)$ as functions of the logarithmic energy for different reconstructions, including the saturated events. Here, $\langle \widehat{\mu}_{\textrm{ML}} \rangle$ and $\sigma[\widehat{\mu}_{\textrm{ML}}](450)$ are estimated from the Gaussian distribution fit of $\widehat{\mu}_{\textrm{ML}}(450)$ histograms. The figure shows that the relative bias for the ADC and ADCProfile reconstruction methods is less than $2\%$ at lower simulated energy and lies within $1\%$ for energies larger than $10^{17.5}$ eV. The value of $\varepsilon(450)$ decreases with primary energy due to the increase in the number of muons triggering more detectors as well as the smaller relative fluctuations at different detectors. The $\varepsilon(450)$ values for the ADC reconstruction are similar to those for the Profile reconstruction. Specifically, $\varepsilon(450)$ for the ADC reconstruction is larger than that for the Profile reconstruction at low energies, but at high energies, it is smaller. The $\varepsilon(450)$ for the ADCProfile method takes smaller values or is equal to those for the ADC and Profile reconstructions. Note that the $\varepsilon(450)$ for the ADCProfile differs by no more than $\sim 2\%$ from that of the ideal counter in the whole energy range considered.
    \paragraph{} Figure \ref{biasvsr} displays the relative bias and relative standard deviation as a function of the distance to the shower axis for iron primaries at an energy of $10^{18.25}$ eV and a zenith angle of 30$\degree$. The UMD reference distance is indicated by a dotted line. The ADC and ADCProfile reconstruction methods exhibit a bias similar to that of an ideal counter at distances nearer to the shower axis. This is because the ADC, which dominates in this region, is more suitable for measuring high values of muon density than the binary mode. The ADC and ADCProfile methods yield a smaller standard deviation compared to the Profile method at distances closer to the shower axis. This is because the ADC dominates in the ADCProfile method at distances less than 450 m from the shower axis, whereas the binary mode dominates at larger distances. The ADCProfile reconstruction method has a relative standard deviation comparable to that of the Profile reconstruction method at distances away from the shower axis. Based on the relative bias and relative standard deviation, the ADCProfile method has the best reconstruction performance.
    \paragraph{} The coverage probability quantifies the quality of the parameter errors. This is the fraction of events in which the confidence interval includes the true value from the simulated MLDF. In the reconstruction, the $1\sigma$ errors of the MLDF normalization $\mu_0$ and the slope parameter $\beta$ are calculated for each event. The parabolic parameter error is obtained directly from the fit procedure \cite{Minuit}. Figure \ref{Cov} displays the coverage of $\widehat{\mu}_{\textrm{ML}}$(450) for the reconstructions of iron primaries at $30\degree$ zenith angle and different energies, using different reconstruction methods. The dotted line at $68\%$ corresponds to the coverage of the $1\sigma$ Gaussian confidence interval. As depicted in the figure, the coverage probabilities of all reconstructions are close to each other and close to the Gaussian reference.
    \paragraph{} To thoroughly examine the impact of saturated events on the reconstruction process, we plot in the top panel of Figure \ref{Nmu_ADC}, the reconstructed value of $\widehat{\mu}_{\textrm{ML}}$(450) for iron primaries with an energy of $10^{18.5}$ eV and a zenith angle of 30$\degree$. The ADC method was used for the reconstruction, and the blue and red histograms indicate the reconstructed candidate and saturated events, respectively. In the bottom panel of the same figure, we have plotted the relative bias and relative standard deviation against energy, after excluding the saturated events for all reconstruction methods. As shown in the figure, the ADC, ADCProfile, and Profile reconstruction methods exhibit better performance and have a relative standard deviation that is closer to that of an ideal counter, particularly at higher energies.
    \paragraph{} In this study, the saturation level in the LG channel of the ADC was set to 1086 muons. To assess the performance of the ADC reconstruction method at various levels of saturation, 10,000 events were reconstructed at lower ($\sim 300$ muons) and higher ($\sim 2500$ muons) saturation levels. Figure \ref{satlev_comp} compares the relative bias and relative standard deviation of $\widehat{\mu}_{\textrm{ML}}$(450) at different levels of ADC saturation. The plot indicates that although a higher ADC saturation levels result in better relative standard deviation, especially at larger energies, this improvement is minimal and not significant compared to the unrealistic saturation level chosen.
    \paragraph{}It is worth noting that the results achieved for iron nuclei at $45^\circ$, as well as for protons at $30^\circ$ and $45^\circ$, are similar to those obtained for iron nuclei at $30^\circ$. In all cases, the ADCProfile reconstruction outperforms both the ADC and Profile reconstructions across the entire energy range analyzed. The relative standard deviation of $\widehat{\mu}_{\textrm{ML}}$(450) for the ADCProfile reconstruction is the closest to that of an ideal counter.
\section{Conclusions}\label{sec:6}
 \paragraph{} The electronics of segmented muon detectors can be designed to operate using two independent modes of acquisition: the binary mode and the ADC mode. Currently, both acquisition modes are available for the underground muon detectors at the Pierre Auger Observatory. In the past, muon detectors at AGASA also had both acquisition modes, which could be used independently or in combination to estimate the density of muons hitting a detector. In this study, a new method was developed to reconstruct the muon lateral distribution function based solely on the ADC mode. The underground muon detectors at the Pierre Auger Observatory were used for this case study. It was shown that the performance of the ADC method is similar to the Profile method developed for the binary mode. Specifically, at lower energies, the Profile method outperforms the ADC-based method slightly, whereas at high energies, the ADC-based method performs slightly better than the Profile method. A second method was also developed in this paper based on both acquisition modes, where the Profile likelihood is considered for muon detectors with relatively low values of the muon density, and the ADC likelihood is considered for high values of the muon density. The performance of the combined method is similar or better than the Profile and ADC methods, depending on the primary energy. The detector resolution corresponding to the standard deviation of the estimated number of muons at 450 m from the shower axis obtained with the combined method is quite close to that corresponding to an ideal counter simulated in the range of energies. The performance of the different reconstruction methods was established from the simulations. The combined method can be used to reconstruct the experimental data, and its consistency with the other methods can be verified.
\begin{acknowledgements}
A.~D.~S.~is member of the Carrera del Investigador Cient\'ifico of CONICET, Argentina. The authors thank the members of the Pierre Auger Collaboration for useful discussions. This work is partially supported by CONICET (PIP 2020-0979).
\end{acknowledgements}
\appendix
\section{Comparison of Gaussian likelihood with compound likelihood}\label{App2}
\begin{figure}[h!]
        \centering
    	\includegraphics[width=0.45\textwidth]{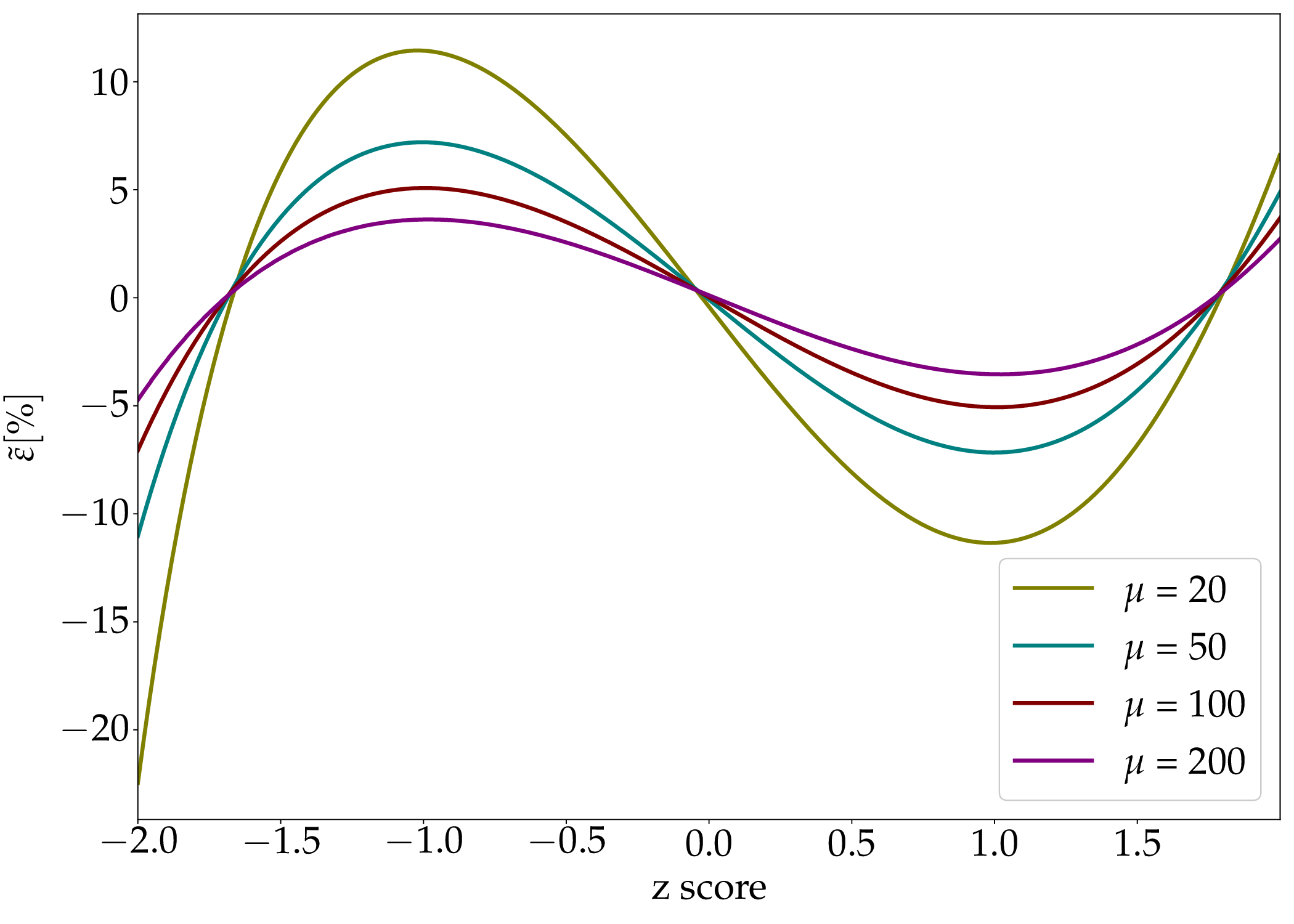}
    	\caption{$\tilde{\varepsilon}$ as a function of the z score for LG channel.}\label{epsz}
    \end{figure}

    \paragraph{}  To compare $f_\textrm{C}(Q|\mu)$ in equation \eqref{convdist} to a Gaussian distribution $f_\textrm{G}(Q|\mu)$ in equation \eqref{Gausseq} more efficiently, we considered a parameter,
    \begin{equation}
    	\tilde{\varepsilon} = 100 \times \left(\dfrac{f_\textrm{G}}{f_\textrm{C}}-1\right) \text{in \%.}
    \end{equation}
    \paragraph{} Figure \ref{epsz} shows the plot of $\tilde{\varepsilon}$ as a function of the z-score for various values of $\mu$. The z-score is defined as,
    \begin{equation}
			\textrm{z-score} = \dfrac{Q-\langle Q \rangle}{\sigma[Q]}.
	\end{equation}	
	\paragraph{} From figure \ref{epsz}, the absolute value of $\tilde{\varepsilon}$ is smaller than 5\% for the LG channel of the ADC, in the $2\, \sigma$ region of the distribution for $\mu=200$. Hence, it is safe to conclude that the $f_C(Q|\mu)$ distribution can be approximated by a Gaussian for values of $\mu$ of the order of or greater than 200.
\section{Study on the effect of the inhibition window on the saturation of the binary mode}\label{App3}
\begin{figure}[h!]
        \centering
    	\includegraphics[width=0.45\textwidth]{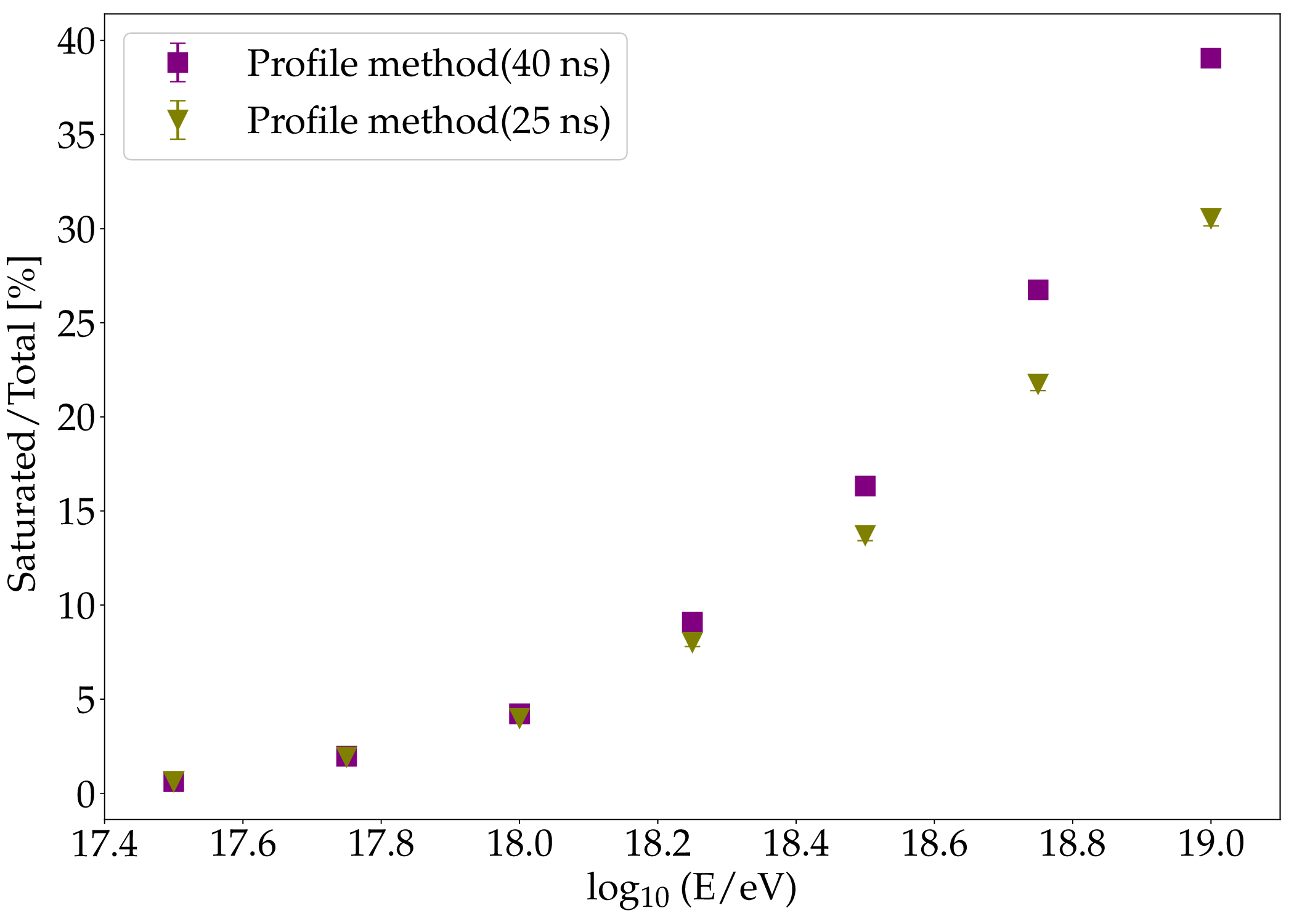}
    	\caption{Fraction of saturated events for iron primaries at 30$\degree$ zenith angle as a function of the logarithm of primary energy for Profile reconstruction method at a 25 ns and at 40 ns inhibition window with error bars. The error bars are smaller than the marker size.}\label{satfrac25}
    \end{figure}

    \paragraph{} The choice of the inhibition window in binary mode has an impact on the number of saturated events. To study this effect in detail, 10,000 events were reconstructed using the Profile reconstruction method with an inhibition window of 25 ns and compared to the 40 ns window used in this analysis. Figure \ref{satfrac25} displays the saturated fraction, with error bars, plotted against the logarithm of energy. As seen from the figure, the choice of inhibition window affects the fraction of saturated events for the Profile reconstruction method. The fraction of saturated events is higher for the Profile method with a 40 ns window and increases rapidly with energy compared to that of the 25 ns window.
\section{Verification of the log-normal character of the sum of \emph{n} log-normal variables}\label{App1}
\begin{figure*}[ht!]
    \centering
    \includegraphics[width=0.9\textwidth]{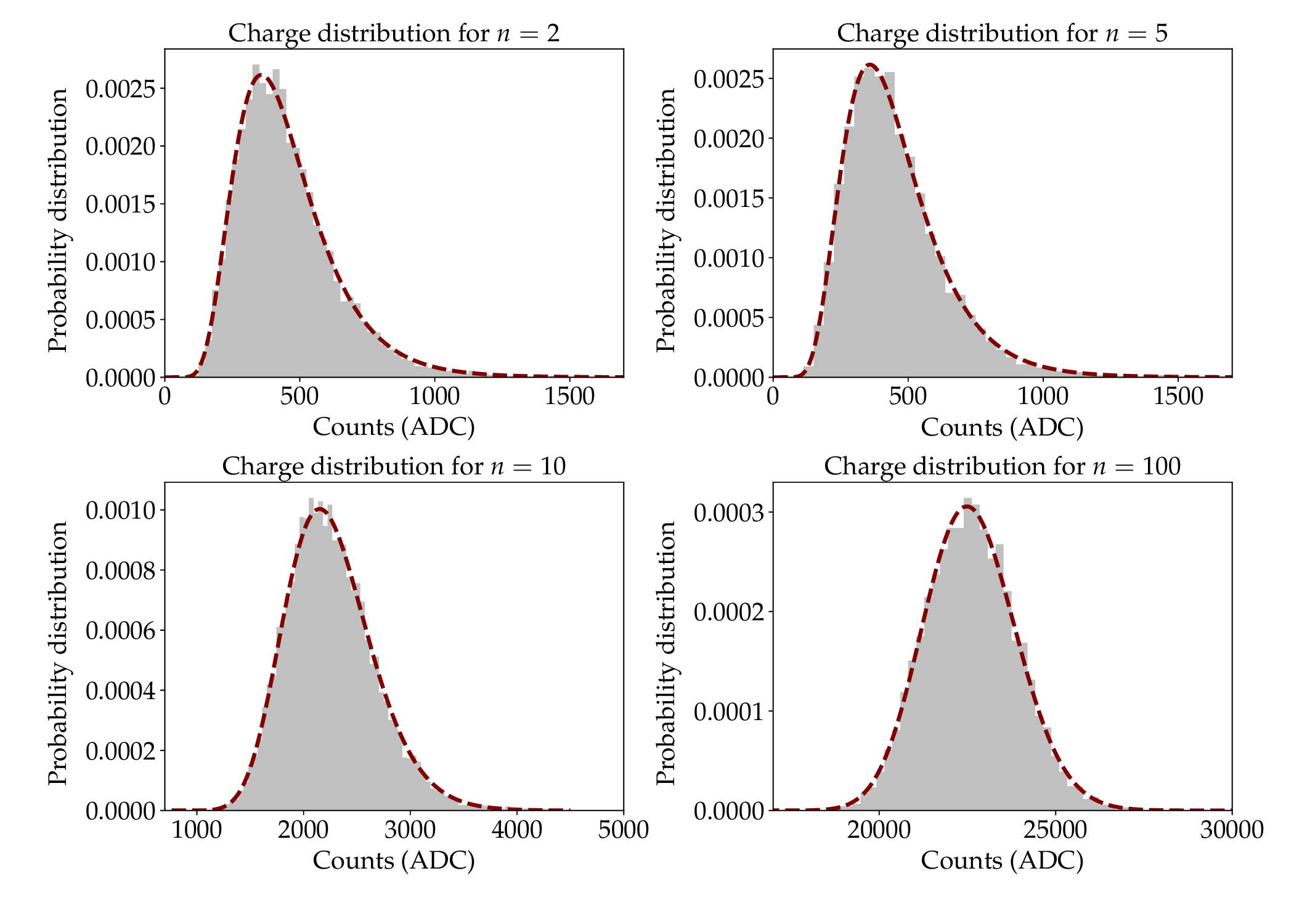}
    \caption{Comparison of the simulated charge distribution with the analytical expression in equation \eqref{Logsum} for the LG channel of the UMDs at the Auger Observatory.} \label{PLn}
    \hfill
    \end{figure*}

    \paragraph{} As shown in equation \eqref{Logsum}, when $\theta^2 \lesssim 1$, the sum of $n$ log-normal terms is expected to exhibit log-normal behavior. Simulations confirmed that $Q$ possesses log-normal characteristics. Figure \ref{PLn} compares histograms of the sum of $n$ random log-normal variables with corresponding log-normal functions (dotted lines) with parameter values $m = 5$ and $\theta = 0.5$ \cite{Botti:19}. 
    \paragraph{} The figure shows that equation \eqref{Logsum} is an excellent approximation for the LG channel. Therefore, when the variance of the distribution is small ($\theta^2 \lesssim 1$), the sum of $n$ log-normal terms behaves similarly to the sum of narrowly distributed random variables, regardless of the value of $n$. The figures reveal that, for smaller $n$ values, the distribution has a prominent tail toward positive total charge values, indicating a larger shape parameter. However, for $n=100$, the tail of the distribution is almost negligible, and the shape parameter is insignificant.

\end{document}